\documentclass[lettersize,journal]{IEEEtran}
\usepackage{cite}
\usepackage{amsmath,amssymb,amsfonts}
\usepackage{graphicx}
\usepackage{xcolor}
\usepackage{hyperref}
\usepackage{mciteplus}
\usepackage{relsize}
\usepackage{array}

\usepackage{textcomp}

\makeatletter
\let\MYcaption\@makecaption
\makeatother
\usepackage{caption}
\usepackage[font=footnotesize]{subcaption}
\makeatletter
\let\@makecaption\MYcaption
\makeatother

\usepackage{microtype}

\usepackage{multirow}
\usepackage{booktabs}
\usepackage{amssymb}
\usepackage[ruled, linesnumbered]{algorithm2e}
\usepackage{booktabs}

\SetKwInput{KwInput}{Input}
\SetKwInput{KwOutput}{Output}
\SetKwInput{KwInit}{Init}

\def\ve#1{{\mathchoice{\mbox{\boldmath$\displaystyle #1$}}%
              {\mbox{\boldmath$\textstyle #1$}}%
              {\mbox{\boldmath$\scriptstyle #1$}}%
              {\mbox{\boldmath$\scriptscriptstyle #1$}}}}

\def\diag{\operatorname*{diag}}
\def\e{\mathrm{e}}

\usepackage{textcomp}
\makeatletter
\def\ps@IEEEtitlepagestyle{%
  \def\@oddfoot{\mycopyrightnotice}%
  \def\@oddhead{\hbox{}\@IEEEheaderstyle\leftmark\hfil\thepage}\relax
  \def\@evenhead{\@IEEEheaderstyle\thepage\hfil\leftmark\hbox{}}\relax
  \def\@evenfoot{}%
}
\def\mycopyrightnotice{%
  \begin{minipage}{\textwidth}
  \centering \scriptsize
  This work has been submitted to the IEEE for possible publication. Copyright may be transferred without notice, after which this version may no longer be accessible.
  \end{minipage}
}
\makeatother

\begin{document}
\title{Velocity-Based Channel Charting with Spatial Distribution Map Matching}

\author{%\hspace{1cm}
Maximilian Stahlke, %~\IEEEmembership{Member,~IEEE},
    George Yammine, %~\IEEEmembership{Member,~IEEE},
	Tobias Feigl, %~\IEEEmembership{Member,~IEEE},
	Bjoern M.\ Eskofier, %~\IEEEmembership{Senior Member,~IEEE},
	Christopher Mutschler %,\IEEEmembership{Member,~IEEE}%\\

\thanks{This work was supported by the Fraunhofer Lighthouse project ``6G SENTINEL'' and by the Federal Ministry of Education and Research of Germany in the programme of ``Souver\"an. Digital. Vernetzt.'' joint project 6G-RIC (16KISK020K).}
\thanks{M.\ Stahlke, G.\ Yammine, T.\ Feigl and C.\ Mutschler are with Fraunhofer IIS, Fraunhofer Institute for Integrated Circuits IIS, Division Positioning and Networks, 90411 Nuremberg, Germany (email: maximilian.stahlke@iis.fraunhofer.de%; george.yammine@iis.fraunhofer.de; tobias.feigl@iis.fraunhofer.de; christopher.mutschler@iis.fraunhofer.de
).}
\thanks{B.\ M.\ Eskofier is with the Department Artificial Intelligence
in Biomedical Engineering (AIBE), Friedrich-Alexander-Universit\"at
Erlangen-N\"urnberg (FAU), 91052 Erlangen, Germany (email: bjoern.eskofier@fau.de).}
}

\maketitle

\begin{abstract}

Fingerprint-based localization improves the positioning performance in challenging, non-line-of-sight (NLoS) dominated indoor environments. However, fingerprinting models require an expensive life-cycle management including recording and labeling of radio signals for the initial training and regularly at environmental changes. Alternatively, channel-charting avoids this labeling effort as it implicitly associates relative coordinates to the recorded radio signals. Then, with reference real-world coordinates (positions) we can use such charts for positioning tasks. However, current channel-charting approaches lag behind fingerprinting in their positioning accuracy and still require reference samples for localization, regular data recording and labeling to keep the models up to date.

Hence, we propose a novel framework that does not require reference positions. We only require information from velocity information, e.g., from pedestrian dead reckoning or odometry to model the channel charts, and topological map information, e.g., a building floor plan, to transform the channel charts into real coordinates.
We evaluate our approach on two different real-world datasets using 5G and distributed single-input/multiple-output system (SIMO) radio systems.
Our experiments show that even with noisy velocity estimates and coarse map information, we achieve similar position accuracies as fingerprinting and do not require ground-truth information.

\end{abstract}


\section{Introduction}
\label{sec:introduction}

\IEEEPARstart{I}{ndoor} positioning serves as a key enabler for various downstream tasks in industrial production, health care or networking \cite{laoudias2018survey}. 
Radio-based localization methods \cite{saeed2019state} are, beside other technologies like camera \cite{wu2018image}, LIDAR \cite{elhousni2020survey} or visible-light-based approaches \cite{rahman2020recent}, one of the most promising technologies for indoor localization \cite{saeed2019state}. 
If there are line-of-sight (LoS) conditions, systems based on angle-of-arrival (AoA) \cite{pang2020aoa, yen20223} or time-of-arrival (ToA) measurements \cite{gifford2020impact} achieve high localization accuracies in the centimeter range. 
However, realistic indoor environments often expose non-line-of-sight (NLoS) signals and multipath, which result in a degradation of the positioning accuracy. 
While error-mitigation methods such as NLoS identification \cite{stahlke2020nlos} or error correction \cite{stahlke2021estimating} may yield robust localization in such environments when redundant access points (AP) are available, fingerprinting-based methods also work in NLoS-dominated areas with few APs~\cite{niitsoo2019deep, stahlke20225g, liu2017toward, de2020csi, widmaier2019towards, singh2021machine, jia2022ttsl, Stahlke2023Uncertainty}. 
However, training a fingeprinting model requires an expensive labeling of channel state information (CSI) measurements with reference positions.
Another disadvantage of fingerprinting is that environmental changes may influence the site-specific fingerprints, necessitating regular updates, which in turn requires updated labeled data~\cite{stahlke20225g, widmaier2019towards}.

To overcome these problems, channel charting~\cite{studer2018channel} exploits CSI of radio systems to model the underlying manifold, which reflects the (local) geometry of the environment. 
The most promising approaches explicitly define the manifold, by means of a distance matrix, to model the radio geometry. 
This can be done either by the CSI data itself~\cite{stahlke2023CC, stephan2023angle, le2021efficient, agostini2020channel, schmidt1986multiple, euchner2022improving} or by physical models of movement, e.g., the distance is directly proportional to time for an agent moving at constant velocity~\cite{stephan2023angle}. 
However, compared to supervised fingerprinting, channel-charting-assisted fingerprinting achieves lower accuracies~\cite{stahlke2023CC} or puts strong assumptions on the agent movement pattern~\cite{stephan2023angle}. 
Furthermore, still few labeled samples are required to exploit the channel chart for localization. 
Given that channel charting degrades similar to environmental changes as fingerprinting, a regular manual labeling process is still inevitable.

In this article, we address all these limitations. 
We investigate how noisy velocity information, e.g., from pedestrian dead reckoning (PDR) or odometry systems, help modeling a channel chart. Since trajectory estimation from such velocity estimations is stable for short time horizons, we can derive a sparse distance matrix and learn the global structure of the radio environment with a Siamese neural network.

We exploit topological map information to learn a transformation to the real-world coordinates and eliminate the need for a ground-truth reference system. 
In contrast to previous work, our map matching algorithm learns the spatial distribution of the channel chart along with the transformation to adjust the orientation of the channel chart. Hence, we only need a rough map representation of the environment, e.g., a floor plan.

We evaluate our algorithm with two different radio systems, a 5G-based radio system and a distributed SIMO system to show the independence to radio topologies and architectures. 
We evaluate the impact of different sources of velocity estimation error on channel charting-based localization performance and show that our algorithm is very robust and therefore applicable to low-cost velocity estimation systems.
We also show that combining different trajectories from independent recordings results in optimal charts.
The latter may enable novel crowdsourcing strategies.
Our results show that we can achieve similar results to supervised fingerprinting, eliminating the need for ground-truth reference systems.

The remainder of this article is structured as follows. Sec.~\ref{section:related_work}
discusses related work. Next, Sec.~\ref{section:Method} provides details about velocity based channel charting algorithm and the adaptive map matching. 
Sec,~\ref{section:exp_setup} describes our experimental setup. The numerical results are presented in Sec.~\ref{section:evaluation} and discussed in Sec.~\ref{section:limitations}. Sec.~\ref{section:conclusion} concludes. 

\section{Related work}\label{section:related_work}

Channel charting~\cite{studer2018channel} models the geometry of the radio environment and supports various tasks such as UE grouping \cite{al2021adaptive}, radio resource management \cite{al2020multipoint}, beamforming \cite{ponnada2021location, ponnada2021best, kazemi2021channel}, pilot assignment \cite{ribeiro2020channel} or localization \cite{taner2023channel, stephan2023angle, stahlke2023CC, lei2019siamese, ferrand2021triplet, deng2021network, zhang2021semi}. Channel charting typically runs in two phases: First, estimating the distances between channel measurements that are proportional to the physical distance and modeling the manifold of the channel information. Second, reducing high-dimensional channel measurements to a 2D vector representation that reflects the coordinates of the radio environment.

The distance metrics are mostly based on the free-space path loss of radio signals~\cite{studer2018channel}, with extensions to make it insensitive to fast-fading effects~\cite{le2021efficient} or grouping of collinear measurements~\cite{agostini2020channel}. 
More advanced approaches extract multipath information with multiple signal identification classification (MUSIC) to cluster multipath components (MPCs) and exploit the path-loss for every MPC~\cite{schmidt1986multiple}. 
To utilize more environmental information, Stahlke et al.~\cite{stahlke2023CC} exploited MPC information from power-delay profiles for time-synchronized high-bandwidth single-input/single-output (SISO) radio systems. 
Stephan et al.~\cite{stephan2023angle} extended their metric for multiple-input/multiple-output (MIMO) radio systems to exploit phase information. However, distance estimates based on radio signals are often noisy due to constraints like collinearity and bandwidth limitations. 
Thus, recent approaches model the proximity of channel measurements by time at high accuracy~\cite{ferrand2020triplet, ferrand2021triplet, rappaport2021improving, ghazvinian2021modality}, as they assume that measurements close in time are also close in space and vice-versa. 
However, this only holds for specific movement patterns, e.g., cars, pedestrians walking on straight lines~\cite{euchner2022improving} or movements at constant velocity~\cite{stephan2023angle}.

As the idea of channel charting is to model the 2D manifold given the high-dimensional channel measurements, dimensionalty reduction is crucial. 
There are non-parametric approaches such as principal component analysis (PCA) \cite{studer2018channel}, Laplacian eigenmaps \cite{ponnada2019out} or Isomap \cite{le2021efficient}. 
However, they are mostly restricted in their ability to model non-linearities or efficiently perform predictions on unseen data. 
Thus, parametric, neural-network-based, approaches are the most employed type of algorithms, e.g., autoencoders \cite{geng2020multipoint, huang2019improving}, Siamese networks \cite{stephan2023angle, stahlke2023CC, lei2019siamese} or triplet-based models \cite{ferrand2020triplet, ferrand2021triplet, rappaport2021improving, euchner2022improving, yassine2022leveraging}. 
In contrast to non-parametric model, they work well on unseen data and their non-linearity allows to model the manifold of the channel measurements.

Channel charting only models the radio environment up to isometries which requires a transformation to use them for localization. This is typically done in a semi-supervised way with few ground-truth samples, which allows to unwrap the local geometry of the channel chart into the global geometry and simultaneously aligns the chart with the real world coordinates~\cite{lei2019siamese, ferrand2021triplet, deng2021network, zhang2021semi}. 
If the channel chart already reflects a global geometry, a simple linear transformation is sufficient, which requires only very few ground truth reference positions~\cite{stephan2023angle, stahlke2023CC}. 
However, fingerprinting, such as channel charting, require regular updates to yield accurate and robust positions when the environment changes~\cite{stahlke20225g}. 
Thus, the effort for regularly recording of only few labeled samples is still demanding. 
To avoid labeled data samples, Ghazvinian et al.~\cite{ghazvinian2021modality} exploit map information to perform an alignment of the channel charting coordinates into the environment. 
They define a map as discrete probability density function (PDF), which represents the distribution of the recorded data. 
They transform into real world coordinates by matching the data distribution of the channel chart with the defined map by optimal transport. 
However, their assumption that the spatial distribution of the channel chart matches the distribution of the map is very strong, as this is only true if data logging is explicitly planned.
This is a contradiction to the mechanism of channel charting, which benefits when many uncontrolled data collectors/participants update the map in a crowdsourcing manner.

\section{Method} \label{section:Method}

We ntroduce our novel approach to using velocity data to create a channel chart in Sec.~\ref{meth:velocity}) and describes how we transform the local channel chart coordinates to the real-world coordinates using coarse map information in Sec.~\ref{meth:map}. Sec.~\ref{meth:pos} describes our positioning pipeline.

\subsection{Velocity-Based Channel Charting}\label{meth:velocity}

Channel charting typically requires a distance matrix that defines the manifold of the data, to obtain the chart of the radio environment. 
We generate a sparse distance matrix based on velocity information available during our data recording to estimate local distances between channel measurements. 
Velocity information can be obtained by many devices, especially in indoor environments such as car production lines or storage centers. 
Robot platforms or industrial cars such as forklifts often implement wheel-based odometry~\cite{chong1997accurate} or visual odometry systems~\cite{scaramuzza2011visual}. 
Also many devices carried by persons, e.g., smartphones, include inertial measurement units to estimate velocity information by PDR algorithms \cite{hou2020pedestrian}. 
If the velocity information is free of errors, it could be integrated over time to determine reference positions. 
However, as both odometry and PDR systems are error prone (i.e., odometry due to slipping wheels, varying tire pressure, and PDR due to, e.g., wrong step-length estimation or drifting IMU sensors) an integration of these errors over time leads to large localization errors for long time horizons. Hence, velocity-estimation systems do not serve as stand-alone localization techniques but rather as assisting techniques to existing positioning systems. 
Here, the relative positions are very accurate and therefore enable distance estimation between consecutive measurements for a certain time window. 
We exploit this fact and create a sparse distance matrix, which is then used for channel charting by a Siamese network.

\subsubsection{Sparse Distance Matrix}

\begin{figure}[t]%
	\centering%
    \input{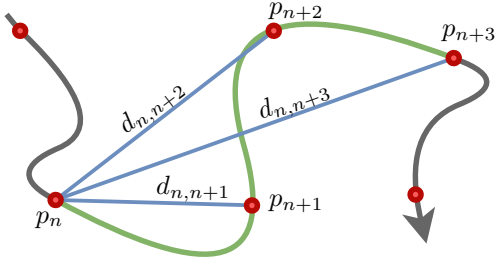}
	\caption{Trajectory of the agent (gray) with consecutive CSI measurements (red dots). The distances $d_{n,n+1}, \ldots, d_{n,n+3}$ are calculated within a window (green) for the positions $p_n, \ldots, p_{n+3}$.}%
	\label{fig:vel_dist}
\end{figure}

While an agent moves along a trajectory in a certain environment, CSI of a radio system is recorded regularly along with the velocity. 
Within a certain window $w$, velocity is integrated over time to estimate distances between the consecutive CSI. 
The distances within a window can therefore be defined as

\begin{align}
    \label{eq:vel_dist}
    d_{n,k} &= \| p_n - p_{k} \|_2^{} \\
            &= \left\| \int_{t=t_n}^{t_k} v(t)\, \mathrm{d}t  \right\|_2^{} \;,
\end{align}

where $t_n$ and $t_k$ are the time of the first measurement and a later measurement within the window, $d_{n,k}$ is the Euclidean distance between the positions $p_n$ and $p_k$ estimated by the velocity $v(t)$. 
As the estimated trajectory drifts over time, we constrain the distance estimation so that $t_k - t_n < w$, while we calculate all distances from the first measurement at $p_n$ to all consecutive measurements until the end of the window. 
Fig.~\ref{fig:vel_dist} visualizes the distance estimation process. 
An agent, i.e., a person or robot, equipped with a PDR or odometry, moves along a trajectory (gray) within the radio environment and regularly records channel measurements (red). Within a certain time window (green) the distances between the position $p_n$ and the consecutive positions $p_{n+1}, \ldots, p_{n+3}$, estimated by the velocity, are calculated. 
Then, the window is moved by one or more positions, dependent on the stride length $s$, and the next distances are calculated until the end of the trajectory. The stride length is only used to lower the amount of data points for radio systems with a high update rate.
This leads to a sparse distance matrix, as only the distances within the windows are calculated.

\subsubsection{Channel Charting}
\label{chap:CC}

Channel charting exploits both the CSI and the distance matrix to unroll the high-dimensional CSI of radio signals into a 2D space reflecting the coordinates of the radio environment. 
There are approaches which solely rely on the distance matrix for dimensionality reduction such as multi-dimensional scaling~\cite{kruskal1964multidimensional} or Sammon's mapping~\cite{sammon1969nonlinear}. 
However, these methods require that the distance matrix fully describes the manifold. In our case, we only have a sparse distance matrix along a trajectory. Further, to enable crowdsourcing and so to enable seemless and continuous channel charting, we combine data from different non-interconnected trajectories, which makes the problem even more difficult. 

Siamese networks can be used to learn a mapping from CSI to 2D coordinates and work well on unseen data points \cite{stephan2023angle, stahlke2023CC, lei2019siamese}.
As they also learn the manifold of the underlying CSI data on a sparse distance matrix we employ a Siamese network to model the manifold of our CSI data.
In addition to manifold learning, a Siamese network also minimizes high-dimensional input data, in our case the CSI of our radio system, to a low-dimensional tensor, in our case 2D coordinates. 
The network minimizes the distance relation \eqref{eq:vel_dist} by using the following loss function 
\begin{equation}
\label{eq:dist_loss}
    \mathcal{L}_\mathsf{d} = \beta \big|d_{n,k}-\|\ve{z}_n^{}-\ve{z}_k^{}\|_2^{} \big| \;,
\end{equation}
where $\ve{z}_n$ and $\ve{z}_k$ are the 2D outputs of the neural network, given the CSI measurements of the radio system, and $\beta$ is a weighting parameter. 
As the error increases over time within a window, distances to measurements further apart from the start of the window are less reliable. 
Thus, we weight our distance estimations linearly within the window in the range $[1,0)$. 

For our Siamese network, we use similar architectures as proposed in \cite{stahlke2023CC}. 
The network consists of 4 convolutional layers for feature extraction, followed by 2 dense layers for discrimination. 
We apply batch normalization for the convolutional layers followed by rectified linear unit (ReLU) activation functions to introduce non-linearities. 
For the last layer, no activation function is used. 
We do not apply local pooling operations between the convolutional layers, as keeping the dimension of time has shown good results in time-series downstream tasks~\cite{ismail2019deep}. 
To reduce the dimensionality, we apply global average pooling right after the last convolutional layer. 
To enhance the receptive field of the convolutional layers with depth, we increased the kernel sizes of the layers. 
As we use CSI from different radio systems, the architectures are adapted to each radio system. 
The parameters of the networks are summarized in Table~\ref{tab:siam_arch}.

\subsection{Adaptive Map Matching}\label{meth:map}
\label{chap:adaptive_map}

Our veloctiy based channel charting can only reflect the radio environment up to isometries. 
Hence, to use channel charting for localization, a transformation from the local channel-charting coordinate system to the real-world coordinate frame is required. To overcome the need for ground-truth positions, Ghazvinian et al.~\cite{ghazvinian2021modality} proposed to use map information of the environment to perform a transformation to the real-world environment. While they require knowledge about spatial distribution of the channel chart, our approach is adaptive and learns the spatial distribution along with the transformation.

\subsubsection{Optimal Transport}

\begin{table}[t]
\caption{Parameters of the Siamese network architectures for the 5G and SIMO radio setups.}
\begin{center}%
\begin{tabular}{cccccc}
\toprule
Layer type & \multicolumn{2}{c}{Channels} & \multicolumn{2}{c}{Kernel size} & Activation \\
\midrule
& 5G & SIMO  & 5G & SIMO & \\
\cmidrule(lr){2-3}\cmidrule(lr){4-5}
Conv. & $8$ & $8$ & $3$ & $3$ & ReLU \\
Conv. & $8$ & $8$ & $5$ & $5$ & ReLU \\
Conv. & $8$ & $8$ & $7$ & $7$ & ReLU \\
Conv. & $8$ & $8$ & $10$ & $10$ & ReLU \\
Avg.\ Pool. & $8\!\!\times\!\!49$ & $32\!\!\times\!\!100$ & --- & --- & --- \\
Fully Con.\ & $200$ & $200$ & --- & --- &  ReLU \\
Fully Con.\ & $2$ & $2$ & --- & ---  & --- \\
\bottomrule
\end{tabular}
\end{center}%
\label{tab:siam_arch}
\end{table}

Their idea is to derive a discrete probability density function from a topological map information, e.g., a floor plan, and match the channel-charting coordinates to this distribution by optimal transport. 
They estimate a transportation matrix $\ve{T} \in \mathbb{R}_+^{s \times t}$, which satisfies the regularized optimal-transport function
\begin{equation}
\label{eq:opt_trans}
    \Lambda( \ve{C},\ve{p}, \ve{q}) = \operatorname*{argmin}_{\ve{T} \in \gamma(\ve{p}, \ve{q})}\langle \ve{T},\ve{C} \rangle - \frac{1}{\lambda}H(\ve{T}) \;,
\end{equation}
where $\langle \cdot, \cdot \rangle$ is the Frobenius inner product, $\ve{p}$ and $\ve{q}$ are probability distributions of samples from the source domain $\Omega_{\mathsf{s}}$ (in our case the distribution of the channel chart) and the target domain $\Omega_{\mathsf{t}}$ (the distribution of the topological map). 
Their joint probability is $\gamma$ and $\ve{C} \in \mathbb{R}_+^{s \times t}$ is a distance matrix, calculated using the Euclidean distance, between the channel-chart coordinates and the map samples. 
It describes the cost to transport probability mass between the channel charting and map domain. 
The solution for \eqref{eq:opt_trans} is 
\begin{equation}
\label{eq:opt_trans_sol}
    \ve{T} = \diag(\ve{a}) \ve{K} \diag(\ve{b}) \;,
\end{equation}
where $\ve{K}=[\e^{-\lambda C_{i,j}}]$, $i = 1,\ldots,s$, $i = 1,\ldots,t$, $\ve{a} \in \mathbb{R}_+^s$ and $\ve{b} \in \mathbb{R}_+^t$ can be calculated by the Sinkhorn--Knopp algorithm~\cite{cuturi2013sinkhorn}:
\begin{align}
   \ve{a} \leftarrow \ve{p} \oslash \ve{K}\ve{b} & \quad \text{and} \quad \ve{b} \leftarrow \ve{q} \oslash \ve{K}^\top \ve{a} \;,
\end{align}
where $\cdot^\top$ is the transpose of the matrix and $\oslash$ is the element-wise division operation. 
The algorithm iteratively estimates \eqref{eq:opt_trans}, while $\lambda$ regularizes the stability of the convergence by controlling the entropy $H(\ve{T})$. 
The higher the entropy $H(\ve{T})$, the faster the convergence but also the less optimal the transport between the probability distributions $\ve{p}$ and $\ve{q}$ is. 
As the algorithm only consists of linear operations it can be differentiated and thus be used as loss function
\begin{equation}
    \mathcal{L}_{\mathsf{m}} = \langle \ve{T}, \ve{C} \rangle 
\end{equation}
to minimize the distribution of the channel chart coordinates and the distribution of the topological map.

\subsubsection{Adaptive Map Distribution}

Ghazvinian et al.~\cite{ghazvinian2021modality} assumed to know the distribution of the source domain $\ve{p}$ (or alternatively assume a uniform distribution within the area of the topological map). 
This only works in scenarios, where the data recording is explicitly performed, such as a person walking along a (meander) path of straight lines in a constant velocity through the entire environment. 
However, in a realistic indoor environment, e.g., an industrial production line, such an uniform distribution is very unlikely. 
There are (temporary) inaccessible areas due to machines or cordoned off areas. In addition, automated guided vehicles (AGVs) may not enter all areas of the environment, as their purpose is to deliver goods between storage areas. However, by assuming a uniform distribution within the environment, the map matching fails, see Sec.~\ref{chap:adaptive_map_eval}. 
Thus, we enhance the approach of Ghazvinian et al.~\cite{ghazvinian2021modality} to learnable probabilities within the discrete map distribution. 
Hence, our approach does not force to match the source-domain distribution to the map but instead penalizes samples that are outside of the possible areas.

\subsubsection{Map Matching}

In contrast to Ghazvinian et al.~\cite{ghazvinian2021modality} that learn the manifold along with the map matching simultaneously, we formulate a two stage optimization approach:
In the first step, we estimate the channel chart, which only reflects the geometry of the environment.
In the second step, we learn a linear transformation, i.e., translation $t_{\theta}$ and rotation $\phi_{\kappa}$, to match the map with the learned channel chart. 

\begin{algorithm}[t]
\caption{Map matching.}\label{alg:map_match}
\KwInput{$X_{\mathsf{s}} = \{x_i\}^{N_\mathsf{s}} \in \Omega_{\mathsf{s}}, Y_{\mathsf{t}} = \{y_i\}^{N_\mathsf{t}} \in \Omega_{\mathsf{t}}, \phi_i$}
\KwOutput{Linear transformation, i.e. $t_{\theta}$, $\phi_{\kappa}$}
\KwInit{$q_{\zeta} = \mathcal{U}, \phi_{\kappa} = \phi_i$}

$t_{\theta} = \text{InitTranslation}(X_{\mathsf{s}}, Y_{\mathsf{t}})$;

\For{$I_\mathsf{iter}$ iterations}
{
    $B = \text{batches}(X_{\mathsf{s}})$\;
    \ForEach{$b$ in $B$}
    {
        $b = \text{linTrans}(b, t_{\theta}, \phi_{\kappa})$\;
        $\mathcal{L}_{\mathsf{m}}$ = $\text{Sinkhorn}(b, Y_{\mathsf{t}}, q_{\zeta})$

        $t_{\theta} \leftarrow t_{\theta} + \eta \frac{\partial \mathcal{L}_{\mathsf{m}}}{\partial t_{\theta}}$\;
        \If{$k > I_\mathsf{wt}$}
        {
            $\phi_{\kappa} \leftarrow \phi_{\kappa} + \eta \frac{\partial \mathcal{L}_{\mathsf{m}}}{\partial \phi_{\kappa}}$\;
        }
        \If{$k > I_\mathsf{wl}$}
        {
            $q_{\zeta} \leftarrow q_{\zeta} + \eta \frac{\partial \mathcal{L}_{\mathsf{m}}}{\partial q_{\zeta}}$\;
        }
    }
}
\end{algorithm}

\begin{figure*}[t!]%
	\begin{subfigure}[h]{1.0\columnwidth}%
        \centering%
		\includegraphics[clip, width=1.0\columnwidth]{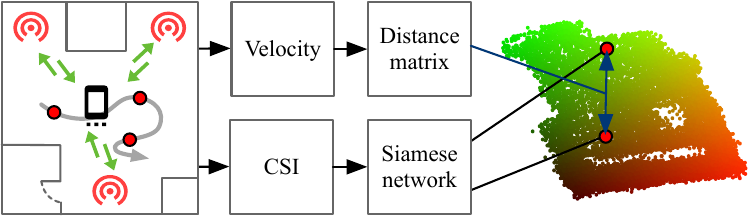}%
        \subcaption{Stage 1: Channel charting.}
        \label{fig:pos_pipe_stage_1}
	\end{subfigure}%
	\begin{subfigure}[h]{1.0\columnwidth}%
        \centering%
		\includegraphics[clip, width=0.85\columnwidth]{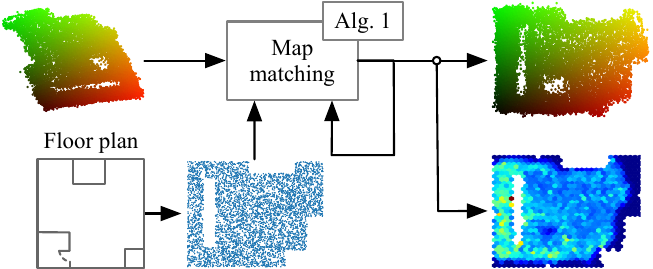}%
        \subcaption{Stage 2: Map matching.}
        \label{fig:pos_pipe_stage_2}
	\end{subfigure}%
	\caption{The two stages of the positioning pipeline. Stage 1 generates a channel chart based on CSI and velocity information. After the channel chart is generated, it only reflects the radio geometry up to isometries. To exploit the channel chart for localization, stage 2 learns a linear transformation to the real world coordinates by provided topological map information.}%
	\label{fig:pos_pipe}
\end{figure*}%

Alg.~\ref{alg:map_match} shows the pseudo-code of our map-matching procedure. The channel chart coordinates of the training data $X_{\mathsf{s}}$ of size $N_s$, the samples of the topological map $Y_{\mathsf{t}}$ of size $N_t$ and the initial rotation $\phi_{i}$ are fed to the algorithm. 
The discrete probability distribution $q_{\zeta}$ of the map is initialized uniformly for all samples. 
In Line 1, the translation between the channel chart $X_{\mathsf{s}}$ and map $Y_{\mathsf{t}}$ is initialized to match their centers of mass. 
Next, our optimization runs for $I_\mathsf{iter}$ epochs on the channel-chart coordinates split into $B$ batches due to memory constraints (Line 3).  
The translation $t_\theta$ and rotation $\phi_{\kappa}$ are applied to the channel chart coordinates (Line 5) and the Sinkhorn distance is estimated (Line 6). 
In Line 7, first the Sinkhorn distance is minimized w.r.t.\ the translation $t_{\theta}$ for $I_\mathsf{wt}$ periods to ensure the translation $t_\theta$ converged before we also start to optimize the rotation $\phi_{\kappa}$.
Once, the parameters of the linear transformation are estimated using the static map, i.e., $I_\mathsf{wl}$, $I_\mathsf{wl} > I_\mathsf{wt}$ periods, the probabilities of the map distribution $q_{\zeta}$ are also optimized to adapt the map to the data distribution. 
We found that, the convergence is very sensitive to the initial rotations $\phi_{i}$, and often ends up in local minima. 
The channel chart may also have inverted $x$- and $y$-axis. 
Thus, we propose to repeat the map-matching algorithm several times and select the transformation parameters with the smallest Sinkhorn distance.

\subsection{Positioning Pipeline}\label{meth:pos}

\begin{figure}[b]%
	\centering%
	\begin{subfigure}{.48\columnwidth}%
		\includegraphics[clip, width=\columnwidth]{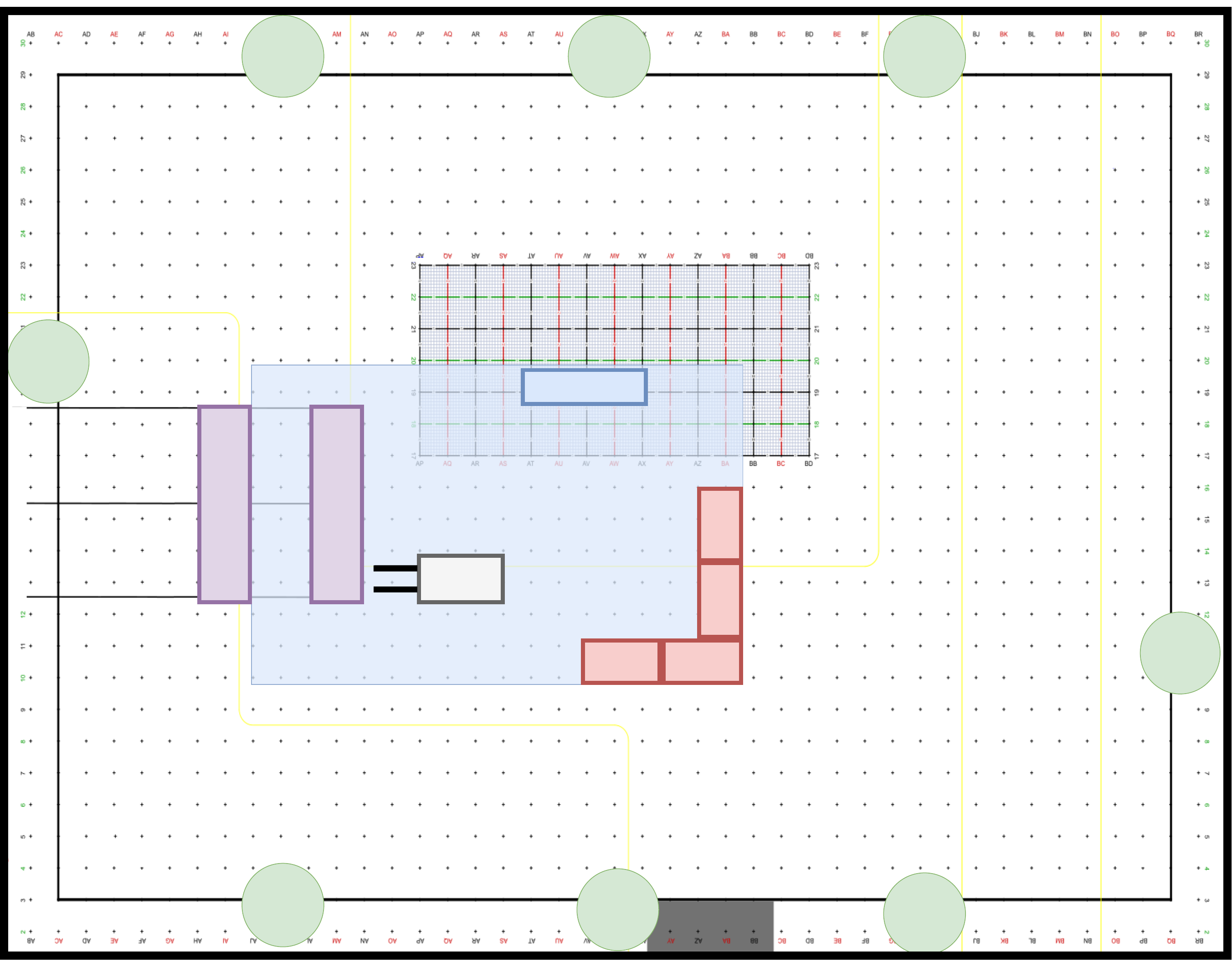}%
	\end{subfigure}%
	\hspace{2pt}%
	\begin{subfigure}{.495\columnwidth}%
		\includegraphics[clip, width=\columnwidth]{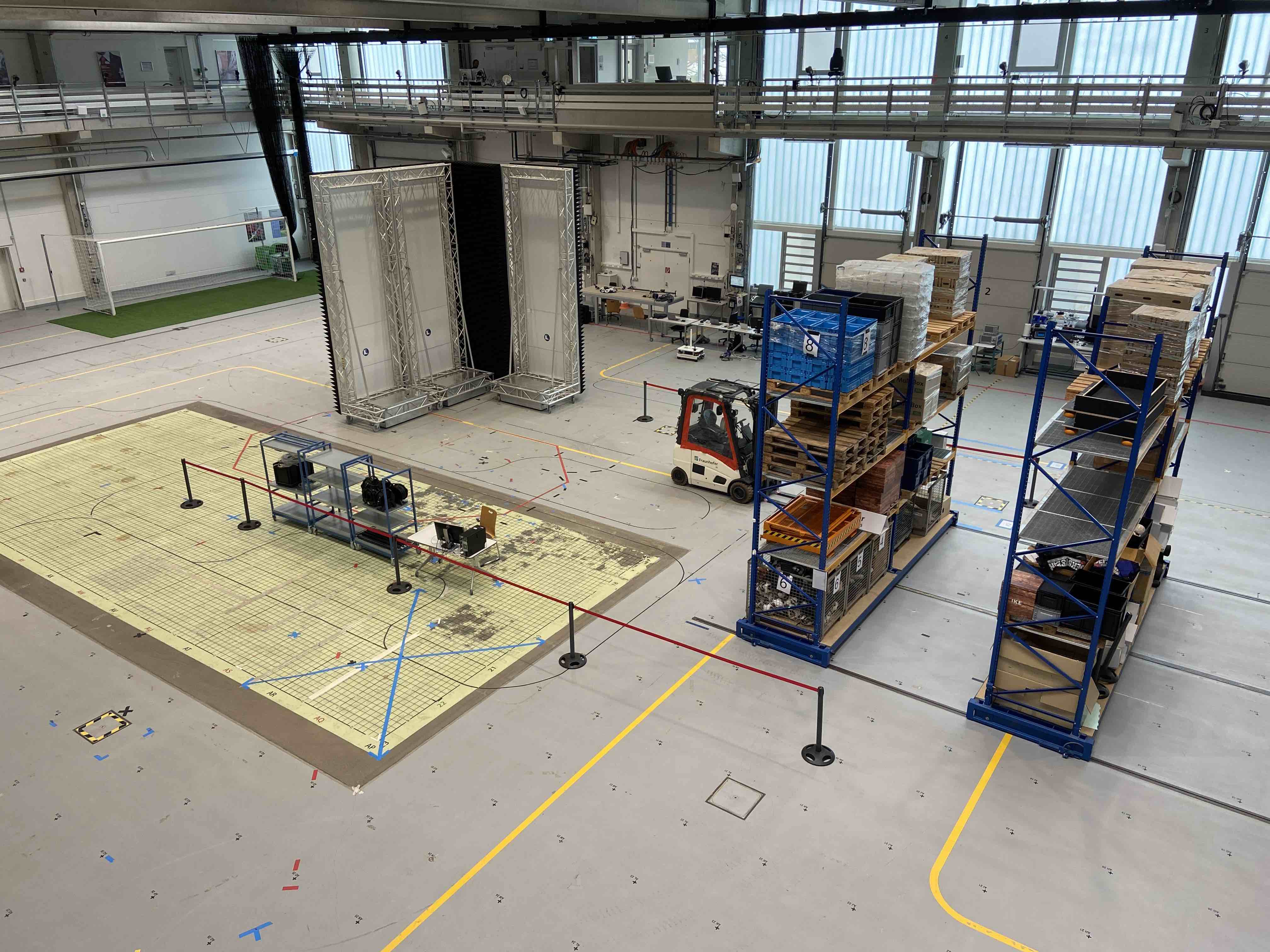}%
	\end{subfigure}%
	\caption{Schematic top view (left) of the environment (right) of the 5G dataset. The red rectangles indicate reflective walls, the green dots are the base stations, blue indicates small shelves and purple the large shelves. The recording area, indicated in blue, has a size of $20\,\mathrm{m} \times 10\,\mathrm{m}$.}%
	\label{fig:env_5G}
\end{figure}%

The final positioning pipeline is shown in Fig.~\ref{fig:pos_pipe} and split into two stages, the first stage (Fig.~\ref{fig:pos_pipe_stage_1}) is used to generate the channel chart, which models the relative positions of the radio unit in the environment, while the second stage (Fig.~\ref{fig:pos_pipe_stage_2}) estimates the transformation from the relative to the global coordinate frame by means of a topological map. 

In the channel-charting stage, shown in Fig.~\ref{fig:pos_pipe_stage_1}, one or more agents move within an environment equipped with an odometry or a PDR system to estimate the velocity. 
While moving, the agent also communicates with a radio system (red) to record CSI (red dots). 
Note that, we assume that both odometry or PDR records are time-synchronized with the CSI records.
After recording, from the velocity information we generate a sparse distance matrix between the CSI measurements, as described in Sec.~\ref{chap:CC}. 
Pairs of CSI measurements are fed to the Siamese network to estimate 2D coordinates for every CSI. 
Eventually, the loss function, described in Eq.~\eqref{eq:dist_loss}, enforces the Siamese network to keep the distances between the measurements and thus models the manifold of the radio signals.
After the training is finished the Siamese network can estimate coordinates from given CSI measurements also on unseen data.

However, the Siamese network can only predict the coordinates up to isometries. 
Hence, to employ it for localization, a linear transformation to the real-world coordinates is necessary. 
To circumvent ground-truth labels, i.e., expensive reference positions, we use our map-matching algorithm as shown in Fig.~\ref{fig:pos_pipe_stage_2}. 
We only require a topological map, e.g., a floor plan. 
The map does not have to cover temporary inaccessible areas, but the areas where no measurements are possible, e.g., outside of the room or in areas with stationary objects, e.g., shelves in a storage hall. 
A discrete PDF is derived from the floor plan and fed to the map-matching algorithm described in Sec.~\ref{chap:adaptive_map}, shown as blue scatters, along with the estimated channel-charting coordinates, shown as the scatters with the color gradient, from the training data distribution. 
The map-matching algorithm learns a linear transformation, which aligns both distributions and simultaneously learns the probabilities, shown as a heatmap in the bottom right of Fig.~\ref{fig:pos_pipe_stage_2}, of the map to adapt to the true data distribution of the training data. 
Finally, a positioning pipeline is created, which estimates the relative coordinates within the channel chart followed by a transformation to the real world coordinates learned by topological map information.

\section{Experimental Setup} \label{section:exp_setup}

To evaluate our approach, we used two different datasets that employ a 5G and a distributed SIMO radio system. 
As the 5G dataset was recorded by a walking person and the SIMO dataset was recorded by a robot platform, the datasets vary in the type of radio signals and in the type of motion.

\subsection{5G Radio System}

\begin{figure}[b]%
	\centering%
	\includegraphics[clip, width=.40\columnwidth]{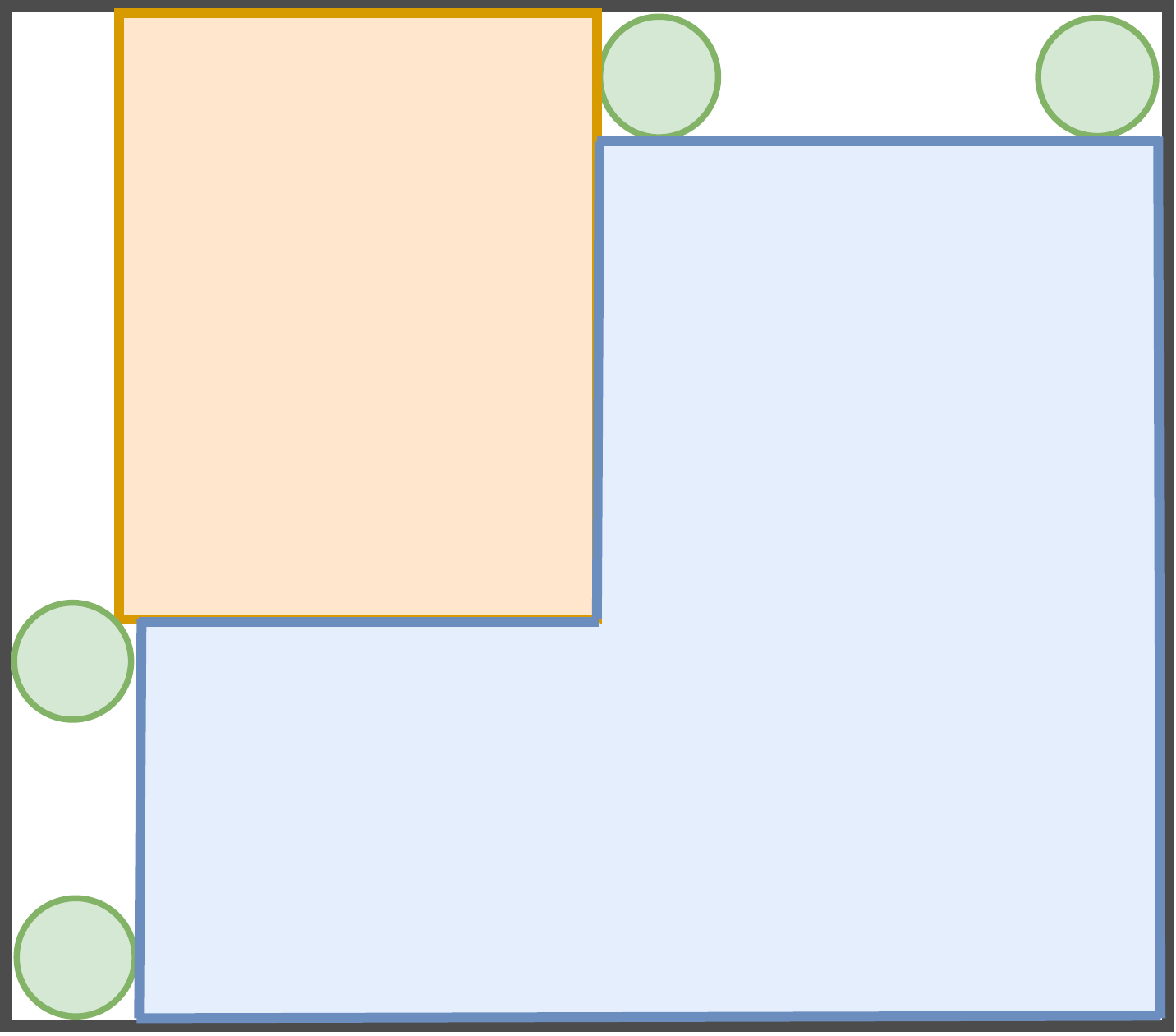}%
	\caption{Schematic top view of the environment of the SIMO dataset \cite{dataDichasusIndustrial}. The orange rectangle indicates a small container room. The recording area, indicated in blue, has a size of $11\,\mathrm{m} \times 13\,\mathrm{m}$.}%
	\label{fig:env_MIMO}
\end{figure}%

For our first experiment, we use a experimental 5G uplink time-difference-of-arrival (TDoA) setup with eight commercial off-the-shelf software-defined-radio base stations (BSs). 
The radio system has a center frequency of 3.75\,GHz with a bandwidth of 100\,MHz, while the BSs are synchronized by a signal generator. 
The data is recorded with a frequency of 100\,Hz. Fig.~\ref{fig:env_5G} shows a schematic sketch of the environment on the left-hand side, and a photo of the real-world setup is shown on the right. 
The base stations (green dots) are placed at the edges of the environment. 
The environment emulates a small industrial setup with large shelves (purple), a working desk (blue) large reflective walls (red) and a forklift (gray). 
The large reflective walls block the signals that impinge on their backside, thus, in the proximity of the walls the majority of the base stations are in NLoS to the transmitter. 
In addition, in-between the large shelves, the signals to the BSs are blocked or distorted. 
We use a mobile phone with a directional antenna as our transmitter.
The mobile phone is carried directly in front of a person, hence shadowing the signals w.r.t.\ the person's point-of-view. 
The statistics of the data recording are shown in Table~\ref{tab:datasets}. 
We recorded two sub-datasets for training with 90{,}417 and 60{,}432 samples respectively and a mean velocity of 0.98\,m/s and 0.87\,m/s respectively. 
Due to the natural movement of a person with various standstill moments, the standard deviation (std.\ dev.) of the velocity is fairly high with 0.45\,m/s and 0.47\,m/s. 
The test trajectory is shorter with only 18{,}256 samples.
\begin{table}[t]
\centering%
\caption{Statistics of the datasets from the radio systems.}%
\begin{tabular}{cccc}
\toprule
Radio system          & Type  & \# Samples & \begin{tabular}[c]{@{}c@{}}Velocity {[}m/s{]}\\ (Mean +- std.\ dev.)\end{tabular} \\ \midrule
\multirow{3}{*}{5G}   & Train & 90.417   & 0.98 +- 0.45                                                               \\ 
                      & Train & 60.432   & 0.87 +- 0.47                                                                 \\ 
                      & Test  & 18.256   & 0.93 +- 0.47                                                                 \\ \midrule
\multirow{2}{*}{SIMO} & Train & 59.137   & 0.28 +- 0.10                                                                 \\  
                      & Test  & 23.478   & 0.25 +- 0.10                                                                 \\ \bottomrule
\end{tabular}

\label{tab:datasets}
\end{table}

\subsection{SIMO Dataset}

For the second experiment, we utilized data from a distributed SIMO radio system \cite{dichasus2021, dataDichasusIndustrial}. 
The orthogonal frequency-domain modulation (OFDM) radio system consists of 4 BSs with $2 \times 4$ antenna arrays each. 
The center frequency is 1.272\,GHz with a bandwidth of 50\,MHz. 
All antennas are synchronized in frequency, time and phase by means of over-the-air synchronization. 
The transmitter is equipped with an omnidirectional antenna. 
Fig.~\ref{fig:env_MIMO} shows the schematic top-view, whereas the arrays (green) are placed at the edges of the ``L-shaped'' recording area (blue) in a research factory campus environment. 
The environment contains a metal container room (orange), which causes NLoS and multipath propagation. 
Data recording is done by a robot platform. 
A training dataset with 59{,}137 samples and a test dataset with 23{,}478 samples are employed as shown in Table \ref{tab:datasets}. 
The velocities are slow compared to the other recordings with only 0.28\,m/s for the training and 0.25\,m/s for the test dataset. The standard deviation of the velocity is only 0.10\,m/s, which indicates a slow and consistent movement.

\subsection{Preprocessing}
\label{chap:preprocess}
To generate characteristic representations of our CSI data for our Siamese network, we use the same preprocessing scheme as proposed in \cite{stahlke2023CC}. 
We use the power-delay profile of the CSI data, since our approach does not rely on phase synchronization. 
We generate a 2D input tensor of dimensions $N_\mathsf{A} \times L_\mathsf{w}$, with $N_\mathsf{A}$ BS and $L_\mathsf{w}$ samples of the CSI. 
For the 5G system, the CSIs are padded by the TDoA to reflect the relative time alignment of the impinging signals. 
Since the CSI data of the SIMO system is already synchronized in time, we do not need to do any additional padding of the CSI.

\begin{figure}[t]%
	\centering%
	\begin{subfigure}{.495\columnwidth}%
		\includegraphics[clip, width=\columnwidth]{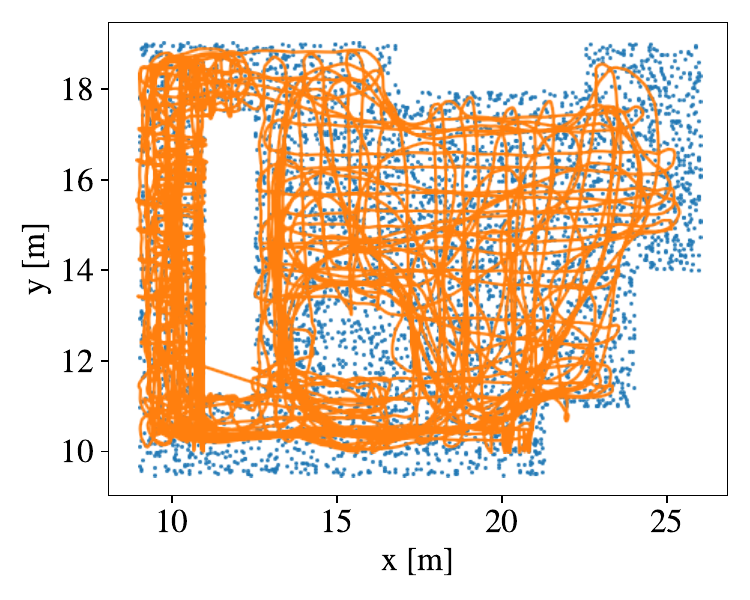}%
	\end{subfigure}%
	\hspace{2pt}%
	\begin{subfigure}{.495\columnwidth}%
		\includegraphics[clip, width=\columnwidth]{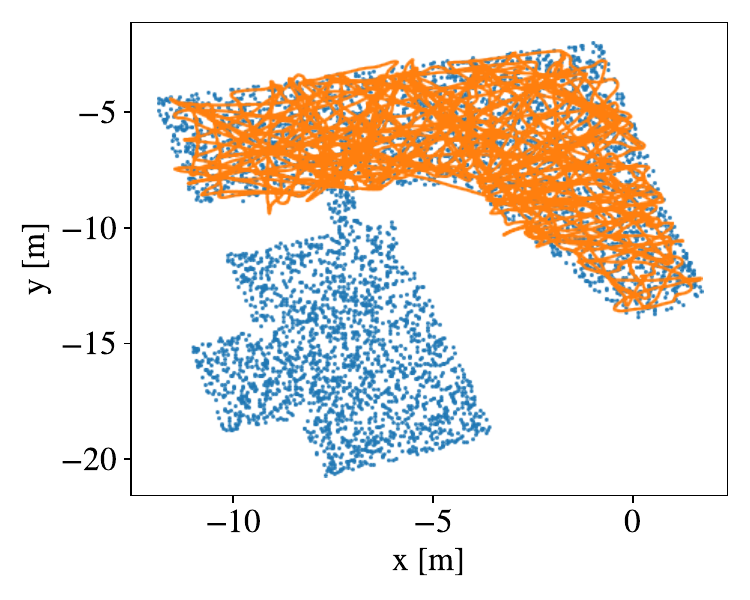}%
	\end{subfigure}%
	\caption{The topological map of the 5G dataset (left) and the SIMO dataset (right) represented as discrete coordinates (blue). The trajectories of the training datasets are shown in orange.}%
	\label{fig:maps}
\end{figure}%

\subsection{Maps} \label{sec:maps}

To learn a transformation from the local channel-chart coordinates, map information is used as described in Sec.~\ref{chap:adaptive_map}. 
The map consists of discrete positions within the area of the topological map and has to provide a unique match of the channel chart within the map. 
Fig.~\ref{fig:maps} shows the maps for the 5G dataset (left) and the SIMO dataset (right). 
The map coordinates are shown in blue, while the orange trajectories show the training datasets. 
The number of samples within the map has to be large enough to cover all areas in the map, while the size is limited to the GPU memory as we have to calculate a distance matrix with the channel-charting coordinates. 
We found experimentally that a size of 5{,}000 samples is large enough for both environments. 
Each of the points require a probability, which describes how likely it is that channel-chart coordinates are placed at this position. 
As we do not have ground-truth reference positions, we are not aware of the spatial density of our channel charts. So, we assign a uniform probability to all the map coordinates.

The 5G map is restricted to the recording area, indicated in blue in Fig.~\ref{fig:env_5G}. 
We consider all static objects in the environment, i.e., the large metal shelves on the left-hand side, the big reflecting walls and the work desks. 
However, the forklift is not included in the map as its position changes regularly and would also neither be included in a floor plan. 
The map of the SIMO dataset is derived from the ground-truth coordinates of the training data distribution, as we have to information about the composition of the environment. 
Beside the area, which covers the training data trajectories, we added an artificial room with no data included. 
Here, we simulate an area in the map the robot cannot access. 
Hence, there is no data available in this area. 
However, the training data, and thus the channel chart, still has a unique match to the map.

\subsection{Velocity Simulation} \label{chap:vel_sim}

Velocity estimation provided by odometry or PDR systems is often error-prone due to slipping wheels, varying tire pressure, wrong step-length estimation or drifting IMU sensors. 
Thus, in this work, we want to investigate the effect of different error sources of the velocity estimation on the channel-charting performance. 
As the recorded datasets do not provide velocities, we estimated the true velocities by the ground-truth labels and added different error sources summarized in Table~\ref{tab:vel_error}.

\begin{figure}[b]%
	\centering%
	\begin{subfigure}{.49\columnwidth}%
		\includegraphics[clip, width=\columnwidth]{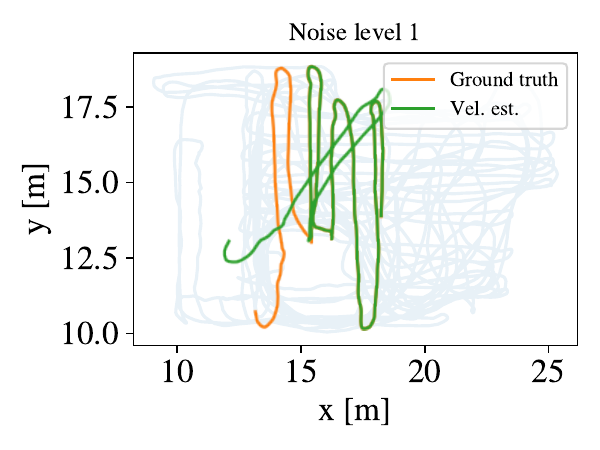}%
	\end{subfigure}%
	\hspace{2pt}%
	\begin{subfigure}{.49\columnwidth}%
		\includegraphics[clip, width=\columnwidth]{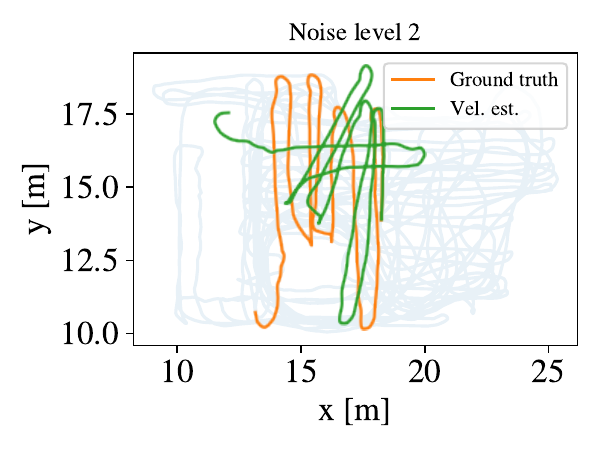}%
	\end{subfigure}%
    %\vspace{2pt}%
    
 	\begin{subfigure}{.49\columnwidth}%
		\includegraphics[clip, width=\columnwidth]{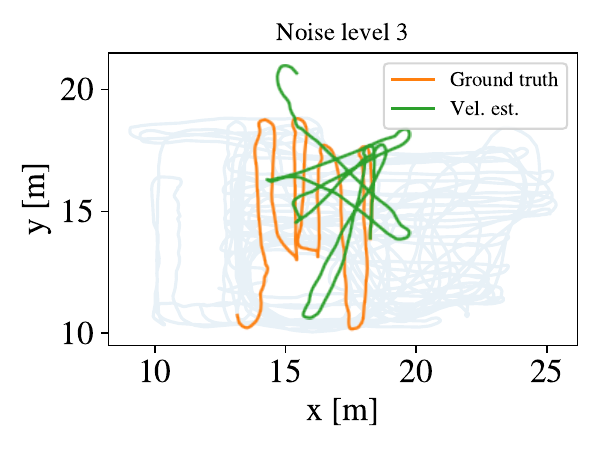}%
	\end{subfigure}%
    \hspace{2pt}%
 	\begin{subfigure}{.49\columnwidth}%
		\includegraphics[clip, width=\columnwidth]{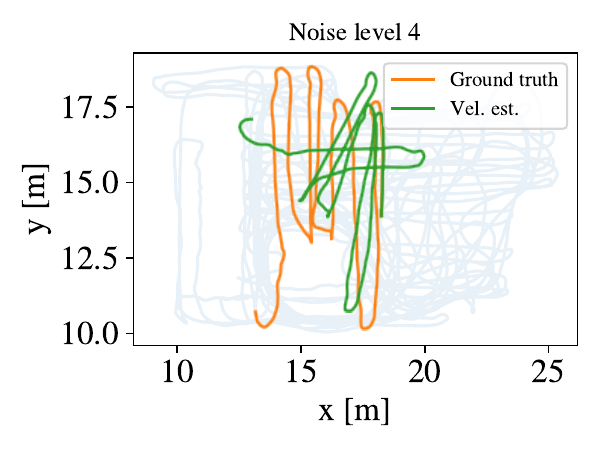}%
	\end{subfigure}%
	\caption{Example trajectories of noise levels 1 to 4. The full trajectory is shown in blue, while a 60\,s window is marked in orange. The trajectory estimated from the noisy velocity is highlighted in green. }%
	\label{fig:vel_noise}
\end{figure}

Three different error sources are considered: a bias in the angular velocity (ang.\ bias), e.g., due to a drifting gyroscope, a bias in the magnitude of the velocity (mag.\ bias), e.g., due to wrong step length estimation and instantaneous rotations (inst. rotations), e.g., due to exceeding the sensitivity limitations of a gyroscope. 
In total, five noise levels are considered with increasing difficulty. 
Example trajectories of noise levels 1 to 4 are depicted in Fig.~\ref{fig:vel_noise}. 
We demonstrate the impact of the noise levels on the trajectory estimation with the 5G training dataset. 
The entire trajectory is shown in blue, a ground-truth window of $60\,\mathrm{s}$ in orange and the estimated trajectory calculated from the noisy velocities in green. 
The first (0) level is noise-free, i.e., the velocity is ideal, while second noise level (1) only contains three different instantaneous rotations, and there are no biases in the angular velocity or magnitude. 
In the third noise level (2), a minor drift of the angular velocity is added along with instantaneous (inst.) rotations, while the angular velocity bias is doubled in noise level four (3). 
The last noise level (4), adds a magnitude bias along with a minor drift of the angular velocity and instantaneous rotations.

\begin{table}[t]
    \caption{Five different noise levels with an bias in the angular velocity (ang.\ bias), a bias in the magnitude of the velocity (mag.\ bias) and instantaneous rotations (inst.\ rotations). The instantaneous rotations are defined by a position within the trajectory, as percentage of the trajectory, and an angle [pos., ang.].}
    \centering
    \begin{tabular}{ccccc}
    \toprule
      \multicolumn{2}{c}{} & \begin{tabular}[c]{@{}c@{}}ang. bias\\ $\mathrm{rad} / \mathrm{s}$ \end{tabular} & \begin{tabular}[c]{@{}c@{}}mag. bias\\ $\mathrm{m} / \mathrm{s}$ \end{tabular} & \begin{tabular}[c]{@{}c@{}}inst. rotations\\ (pos / ang.)\end{tabular} \\ \midrule
      \multirow{6}{*}{\begin{tabular}[c]{@{}c@{}} Noise \\ level\end{tabular}} & 0 & - & -  & -  \\ %\cmidrule{2-5}
       \rule{0pt}{8pt} & 1 & - & -  & $[5, \frac{\pi}{4}]$, $[20, -\frac{\pi}{8}]$, $[60, -\frac{\pi}{6}]$ \\ %\cmidrule{2-5}
       \rule{0pt}{8pt} & 2 & $\frac{\pi}{200}$ & -  & $[5, \frac{\pi}{4}]$, $[40, -\frac{\pi}{8}]$, $[60, -\frac{\pi}{6}]$ \\ %\cmidrule{2-5}
       \rule{0pt}{8pt} & 3 & $\frac{\pi}{100}$ & -  & $[5, \frac{\pi}{4}]$, $[40, -\frac{\pi}{8}]$, $[60, -\frac{\pi}{6}]$ \\ %\cmidrule{2-5}
       \rule{0pt}{8pt} & 4 & $\frac{\pi}{200}$ & $-0.1$ & $[5, \frac{\pi}{4}]$, $[40, -\frac{\pi}{8}]$, $[60, -\frac{\pi}{6}]$ \\ 
         \bottomrule
    \end{tabular}
    \label{tab:vel_error}
\end{table}

\section{Evaluation} \label{section:evaluation}

In the following, we first evaluate the velocity-based channel charting with the adaptive map matching (Sec.~\ref{chap:adaptive_map_eval}), followed by an ablation study on the effects of varying window sizes and velocity noise levels (Sec.~\ref{chap:window_size} and the influence of the numbers of BS (Sec.~\ref{chap:numbers_BS}).
All channel charts are trained for 10{,}000 epochs to ensure the convergence of the networks. For the training of the 5G models, we used both training datasets, described in Table~\ref{tab:datasets}, while we only have one training dataset in the SIMO case. 
For the distance matrix generation, we used a stride length $s=10$ for the 5G datasets and a stride length $s=2$ for the SIMO dataset. 
We employ the training data to optimize the channel chart and to learn the linear transformation with the map-matching algorithm. 
The evaluations are done on separate test datasets on different trajectories to investigate the generalization to unseen data.

\begin{table}[t]
    \centering
    \setlength\tabcolsep{5pt}
    \caption{ Results (CE90) of the map matching algorithms.}
    \begin{tabular}{cccccc}
    \toprule
        Type & FP & Least sq. & \textbf{Ours} & Static map & Comb. \\
        \midrule
        5G & 0.90 & 0.92 & 1.16 & 1.71 & 3.43 \\
        SIMO & 0.56 & 0.71 & 0.90 & 9.65 & 11.08 \\ 
    \bottomrule
    \end{tabular}
    \label{tab:res_map_matching}
\end{table}

\subsection{Velocity based Channel Charting}
\label{chap:adaptive_map_eval}

In this evaluation we investigate the velocity-based channel charting under realistic conditions, i.e., velocity noise level 3, along with the map matching algorithm given coarse, topological map information described in Sec.~\ref{sec:maps}. For the results, we use a window size of 15\,s for the 5G dataset and a window size of 30\,s for the SIMO dataset. 
For all experiments, we employed $I_\mathsf{iter} = 150$ iterations for training, with $I_\mathsf{wt} = 50$ and $I_\mathsf{wl} = 100$ for the warm-up periods and $\lambda = 30$. 
The batch size has to be large enough to reflect the data distribution in the environment. 
We found that a batch size of 3{,}000 samples is sufficiently large for our environments while still matching our memory constraints.
We repeated the map-matching algorithm 60 times, with 20 different equidistant start rotations in the range of $[0,2\pi)$ and inversions of the $x$- and $y$-axis, and selected the transformation with the smallest Sinkhorn distance. 

The results are shown in Table~\ref{tab:res_map_matching}. 
We use three baselines to compare our proposed approach (see \textbf{Ours}). 
We directly compare our velocity-based channel-charting approach to supervised fingerprinting (FP). As fingerprinting model, we used the same model as for the Siamese network and optimized the output directly to the ground-truth coordinates of the CSI values by means of the Euclidean-distance loss as proposed in \cite{stahlke20225g}. As upper bound for the map transformation, we also provide the results with a linear transformation estimated with the ground truth of the training data via least-squares optimization (Least sq.). 
The third baseline (Static map) is similar to our map matching algorithm, however, we do not learn the probabilities of the map along with the linear transformation. 
This is achieved by setting $I_\mathsf{wl} > I_\mathsf{iter}$. 
The fourth baseline (Comb.) follows the idea of \cite{ghazvinian2021modality} by learning the manifold of the data simultaneously to the map matching. 
We employ the same Siamese networks, described in Sec.~\ref{chap:CC}, but use the combined loss ${\mathcal{L}_\mathsf{comb} = \mathcal{L}_\mathsf{m} + \mathcal{L}_\mathsf{d}}$ to optimize the neural network.

\begin{figure}[t!]%
	\centering%
	\begin{subfigure}{.49\columnwidth}%
		\includegraphics[clip, width=\columnwidth]{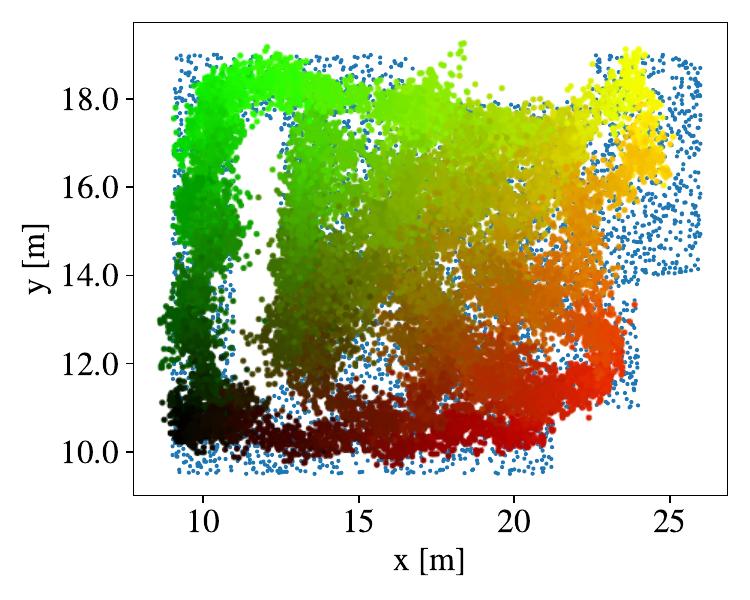}%
	\end{subfigure}%
	\hspace{2pt}%
	\begin{subfigure}{.49\columnwidth}%
		\includegraphics[clip, width=\columnwidth]{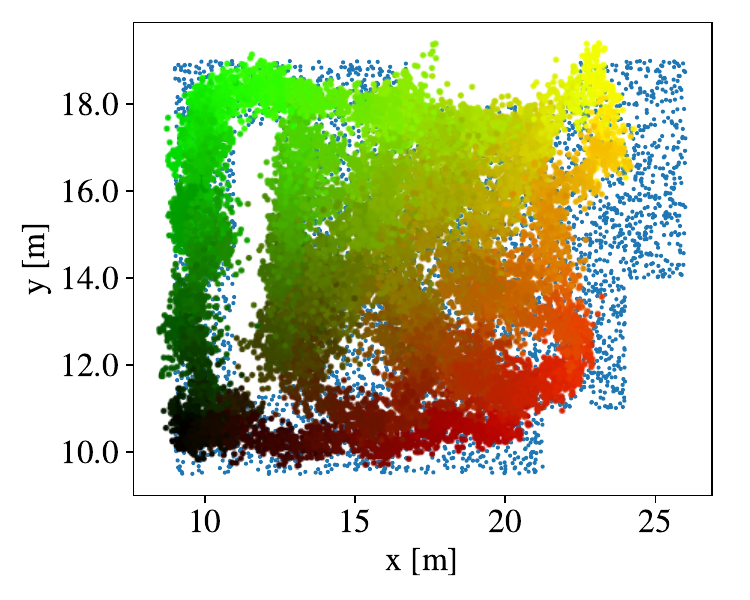}%
	\end{subfigure}%
    %\vspace{2pt}%
    
 	\begin{subfigure}{.49\columnwidth}%
		\includegraphics[clip, width=\columnwidth]{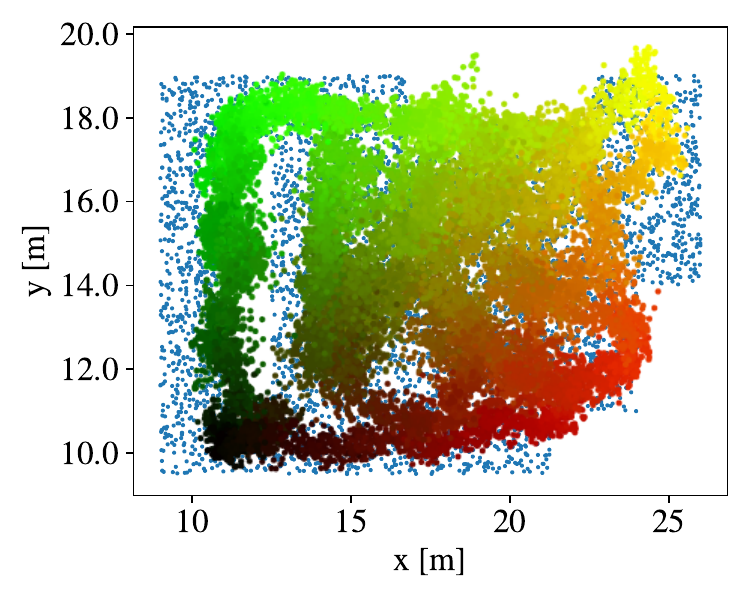}%
	\end{subfigure}%
    \hspace{2pt}%
 	\begin{subfigure}{.49\columnwidth}%
		\includegraphics[clip, width=\columnwidth]{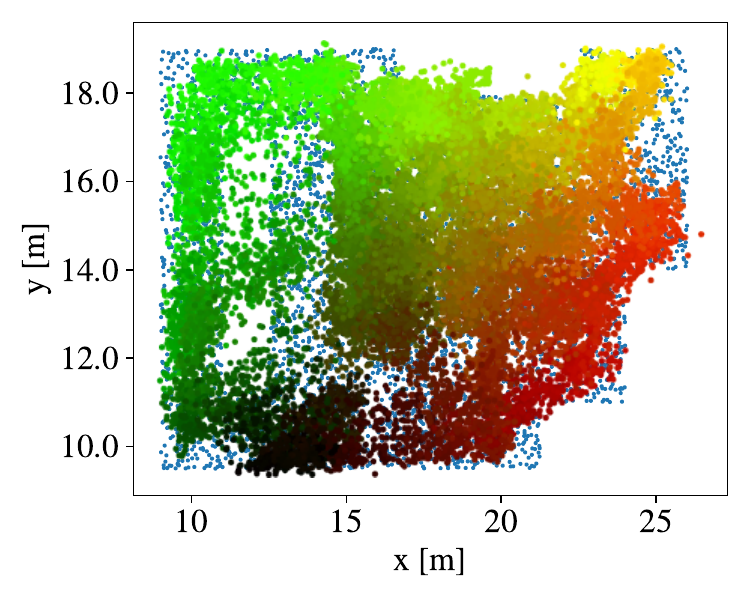}%
	\end{subfigure}%
	\caption{Results of the map matching algorithms for the 5G dataset. The plot top left shows the least squares approach, top right shows the map matching with a trainable map distribution, bottom left the map matching with a static map and bottom right the combined approach. The blue dots show the samples of the map distribution and the dots with the color gradient the channel chart. }
	\label{fig:res_5G_cc_map}
\end{figure}%
\begin{figure}[b]%
	\centering%
	\begin{subfigure}{.495\columnwidth}%
		\includegraphics[clip, width=\columnwidth]{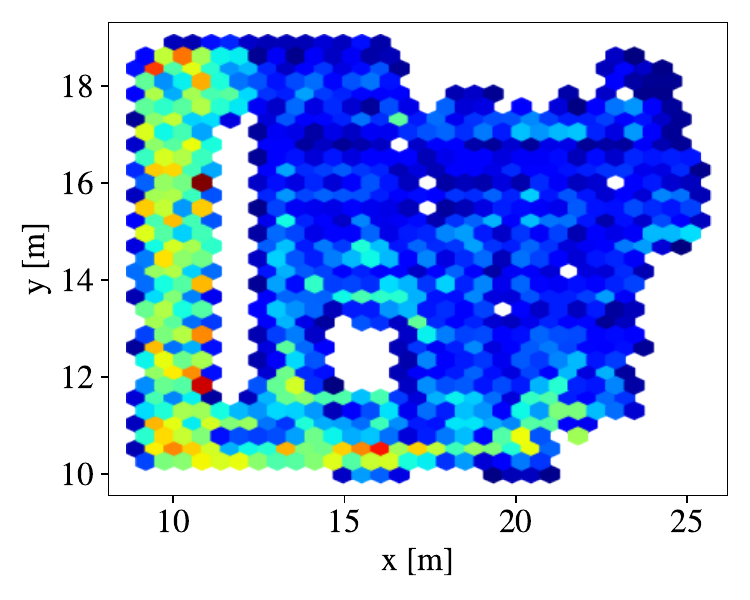}%
	\end{subfigure}%
	\hspace{2pt}%
	\begin{subfigure}{.495\columnwidth}%
		\includegraphics[clip, width=\columnwidth]{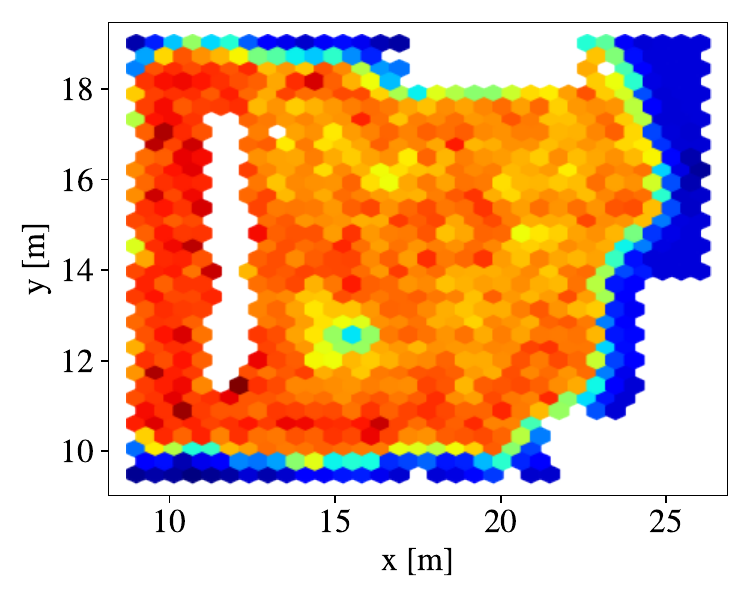}%
	\end{subfigure}%
	\caption{Distribution of the data of the 5G dataset within the environment is shown left, while the learned map is shown on the right. The color indicates the probability, while blue is low and red is high. }%
	\label{fig:5G_data_map_prob}
\end{figure}%

The results for the 5G setup show that our velocity-based channel charting, with a CE90 of 1.16\,m, is close to the FP results, with a CE90 of 0.90\,m, with only a difference of 0.26\,m. Due to the fast velocity of the agent in the 5G dataset, we could employ a small window size of only 15\,s, which leads to only small errors in the distance estimation and thus a high quality of the channel chart. The error arises mainly from the linear transformation to the real world coordinates, as the least squares transformation (Least sq.) can achieve a CE90 of only 0.92\,m, similar to supervised fingerprinting. Fig.~\ref{fig:res_5G_cc_map} shows the map matching of the 5G dataset for the least-squares approach (top left), our approach (top right), the static-map approach (bottom left) and the combined approach (bottom right). 
Our approach has a similar match of the channel chart on the map compared to the least-squares optimization. 
There is only a small difference in the rotation, which already leads to a increased error of 0.24\,m. 
The static-map approach has a higher error with a CE90 of 1.71\,m due to a wrong horizontal alignment, as the channel chart is shifted to the right-hand side. 
This is due to the simplified assumption that the data has a uniform distribution within the map and the lack to adapt the probability mass within the environment. 
Fig.~\ref{fig:5G_data_map_prob}, on the left-hand side, shows the distribution of the positions of the training dataset. It can clearly be seen that on the left-hand side at the large shelf, the data density is much higher compared to the area on the right-hand side. Thus, the channel-charting coordinates are pushed towards the right-hand side for a better overlap of the probability masses, which leads to an error in the translation. 
The combined approach has a similar problem with an even higher CE90 of 3.43\,m. 
While the channel chart coordinates match the area of the map well, the Siamese network tries to match the uniform distribution of the map. 
Hence, points of the left-hand side are pushed towards the right-hand side to match the map distribution. 
Instead, our approach learns the distribution of the probability mass within the map, shown in Fig.~\ref{fig:5G_data_map_prob} on the right-hand side, where blue means low probability and red high. 
We clearly see that the probability mass adapted well to the distribution of the data, shown on the left. 
All areas outside of the area of the training data are assigned a low probability, while the area on the left-hand side has a higher probability like in the data distribution.

\begin{figure}[t]%
	\centering%
	\begin{subfigure}{.49\columnwidth}%
		\includegraphics[clip, width=\columnwidth]{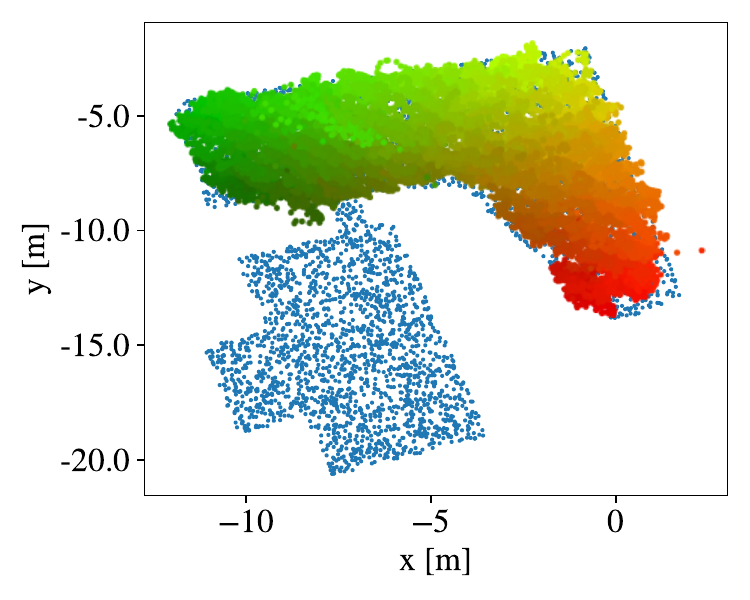}%
	\end{subfigure}%
	\hspace{2pt}%
	\begin{subfigure}{.49\columnwidth}%
		\includegraphics[clip, width=\columnwidth]{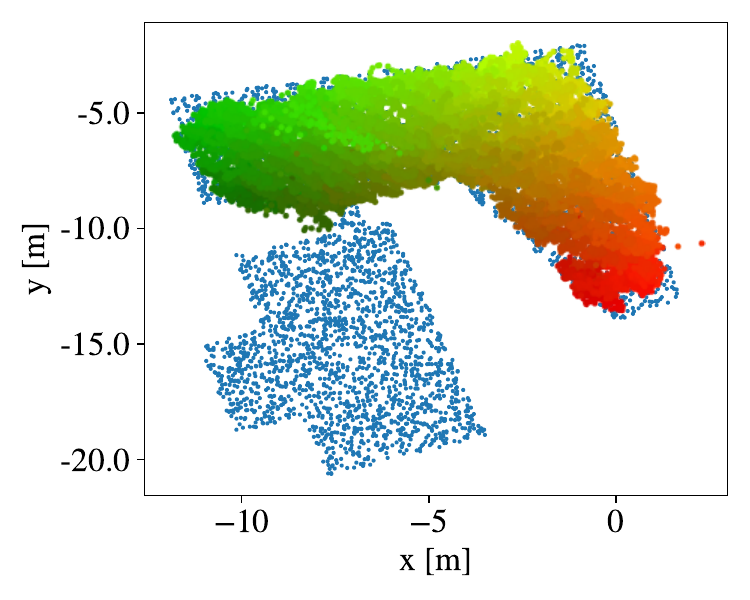}%
	\end{subfigure}%
    %\vspace{2pt}%
    
 	\begin{subfigure}{.49\columnwidth}%
		\includegraphics[clip, width=\columnwidth]{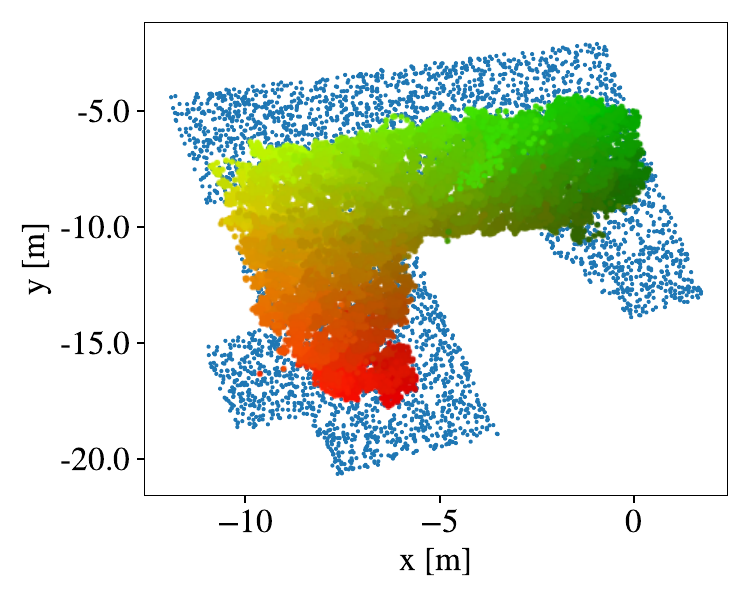}%
	\end{subfigure}%
    \hspace{2pt}%
 	\begin{subfigure}{.49\columnwidth}%
		\includegraphics[clip, width=\columnwidth]{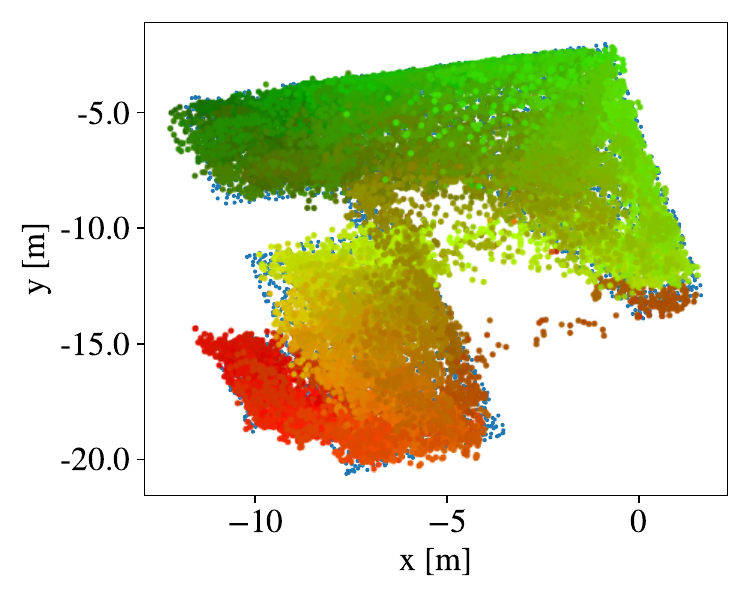}%
	\end{subfigure}%
	\caption{Results of the map matching algorithms for the SIMO dataset. The top left plot shows the least squares approach, the top right plot shows the map matching with a trainable map distribution, the bottom left plot shows the map matching with a static map and the bottom right plot shows  the combined approach. The blue dots show the samples of the map distribution and the dots with the color gradient the channel chart. }
	\label{fig:res_MIMO_cc_map}
\end{figure}%

\begin{figure}[t]%
	\centering%
	\begin{subfigure}{.495\columnwidth}%
		\includegraphics[clip, width=\columnwidth]{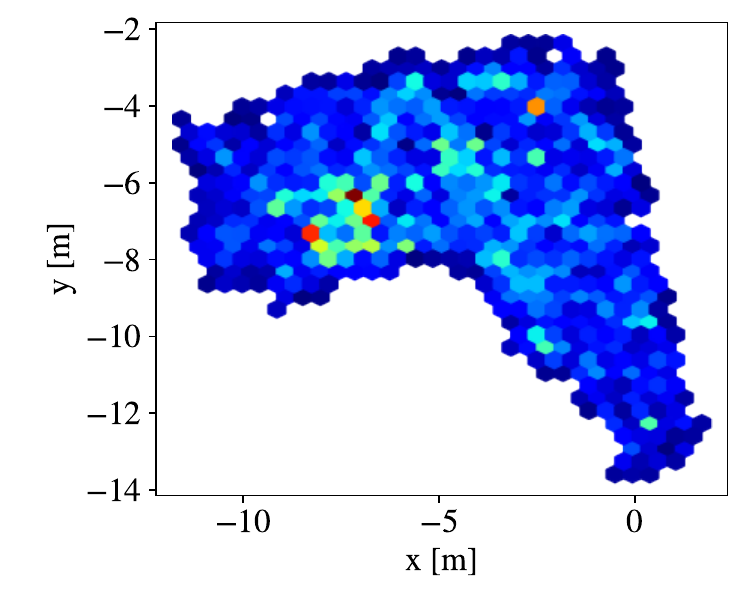}%
	\end{subfigure}%
	\hspace{2pt}%
	\begin{subfigure}{.495\columnwidth}%
		\includegraphics[clip, width=\columnwidth]{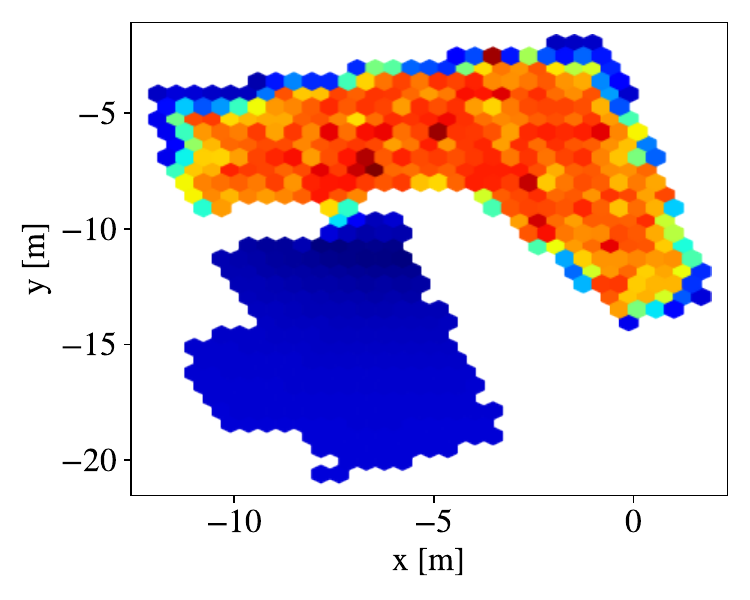}%
	\end{subfigure}%
	\caption{Distribution of the data of the SIMO dataset within the environment is shown left, while the learned map is shown on the right. The color indicates the probability (blue is low and red is high). }%
	\label{fig:SIMO_data_map_prob}
\end{figure}

In the SIMO dataset, our approach achieves only a CE90 of 0.90\,m, which is lower compared to supervised fingerprinting (FP) with a CE90 of 0.56. Due to the slow movement speed of the robot, the window size, i.e., 30\,s, has to be larger to achieve a stable channel chart. More details about the effect on the window size are investigated in Sec.~\ref{chap:window_size}. As the errors of the noisy velocity estimation increases in time, the distance estimations are more erroneous for larger window sizes and hence also the channel chart. Thus, also the channel chart with a optimal least squares can only achieve a CE90 of 0.71. However, our approach is close to the optimal linear transformation with only an difference of 0.19\,m.
The results are visualized in Fig.~\ref{fig:res_MIMO_cc_map}. The least-squares approach is shown top left, our approach top right, the static-map approach bottom left and the combined baseline bottom right. 
Our method has a small error in the rotation compared to the least-squares approach, while the static-map method fails completely. 
As we evaluate our map-matching algorithm with various starting parameters and select the best transformation w.r.t.\ the final Sinkhorn distance, the lowest Sinkhorn distance to a uniform distribution within the map leads to a wrong alignment and thus to a high localization error. 
Also the combined approach achieved a poor localization performance with a CE90 of 11.08\,m. 
The Siamese network matches the map, despite that in the room area, which we artificially added to the map, no data was recorded. 
While the local spatial consistency is still good, i.e., the color gradient is still visible, the Siamese network assign coordinates to the room, which is not covered by the data distribution. 
Our approach, solves this problem by learning that the room does not contain any data points, shown in Fig.~\ref{fig:SIMO_data_map_prob}. 
The left plot shows the distribution of the data, while the right-hand side shows the distribution of the probability mass within the map. 
Our algorithm is able to align the channel chart at the correct position, while learning that the room is not covered by the training data. 
On top of that, this solution ended up with the lowest Sinkhorn distance, as every other alignment of the channel chart would lead to channel-chart coordinates outside of the map area and thus to a higher Sinkhorn distance.

\subsection{Effects of Window Size vs.\ Noise Level} \label{chap:window_size}

In this evaluation, the quality of channel charting is investigated with erroneous velocity estimations and varying window sizes.
We evaluate channel charting isolated from the map matching and use the ground-truth data for the linear transformation to the real-world coordinates.
The algorithm is evaluated for the five different noise levels described in Sec.~\ref{chap:vel_sim}, with four different windows sizes for 5\,s, 15\,s, 30\,s and 60\,s. 
For the training of the 5G setup, we used 8 BSs and for the SIMO setup, 3 BSs were employed. 

The results summarized in Table~\ref{tab:res_window_vs_drift} clearly show that the window size plays a crucial role in modeling the manifold of the CSI data. 
Velocity-based channel charting fails for both datasets with a window size of 5\,s. 
While the 5G dataset can achieve good results for window sizes of 15\,s and larger, the SIMO dataset needs at least a window of 30\,s to achieve a stable result.
We think that the needed window sizes correlate with the velocity of the agent.
In the case of the 5G dataset, the agent had a mean velocity of about 0.95\,m/s, while the mean velocity was much lower in the SIMO dataset with only 0.28\,m/s.
Hence, the distances estimated by the velocities were only small in the SIMO dataset compared to the 5G dataset, which means that the Siamese network was not able to model a globally valid channel chart only based on local distances.

\begin{table}[t]
\caption{Results of the channel charting for different noise levels and window sizes for the 5G and SIMO dataset.}
\centering
\begin{tabular}{ccccccc}
\toprule

\multirow{2}{*}{\begin{tabular}{@{}c@{}}Radio \\ system \end{tabular}} & \multicolumn{1}{c}{\multirow{2}{*}{\begin{tabular}{@{}c@{}}Window \\ size [s] \end{tabular}}} & \multicolumn{5}{c}{Noise level / Error (CE90 in m)} \\ \cmidrule{3-7}
& & 0 & 1 & 2 & 3 & 4 \\ \midrule
\multirow{4}{*}{5G} & 5 & 1.309 & 6.504 & 6.526 & 6.396 & 1.660 \\ %\cmidrule{2-7} 
& 15 & 0.912 & 0.886 & 0.856 & 0.892 & 0.927 \\ %\cmidrule{2-7} 
& 30 & 0.870 & 0.901 & 0.876 & 0.961 & 0.935 \\ %\cmidrule{2-7} 
& 60 & 0.871 & 0.841 & 0.947 & 1.112 & 0.968 \\ \midrule \midrule 

\multirow{4}{*}{SIMO} & 5 & 4.562 & 3.397 & 5.064 & 4.733 & 4.754 \\ %\cmidrule{2-7} 
& 15 & 3.235 & 0.675 & 3.660 & 2.863 & 2.627 \\ %\cmidrule{2-7} 
& 30 & 0.543 & 0.551 & 0.60  & 0.696 & 0.699 \\ %\cmidrule{2-7} 
& 60 & 0.537 & 0.566 & 0.70 & 0.951 & 0.895  \\ \bottomrule

\end{tabular}
\label{tab:res_window_vs_drift}
\end{table}

\begin{figure}[b]%
	\centering%
	\begin{subfigure}{.49\columnwidth}%
		\includegraphics[clip, width=\columnwidth]{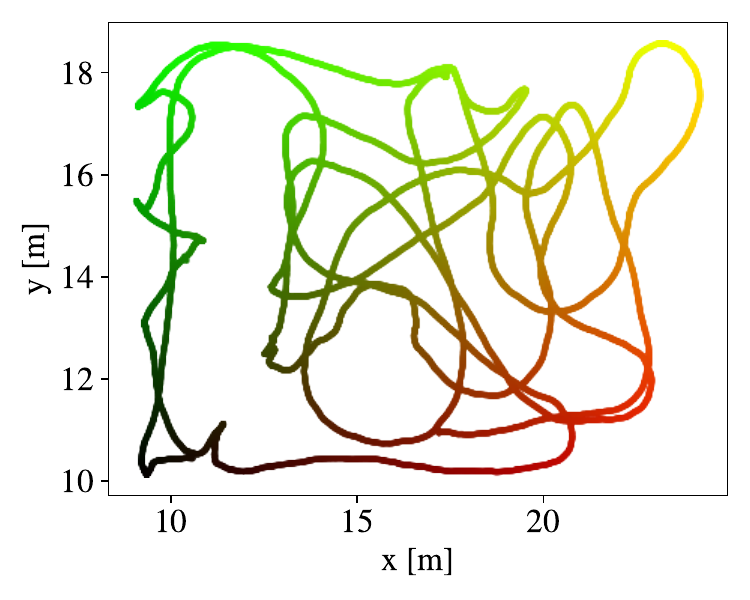}%
	\end{subfigure}%
	\hspace{2pt}%
	\begin{subfigure}{.49\columnwidth}%
		\includegraphics[clip, width=\columnwidth]{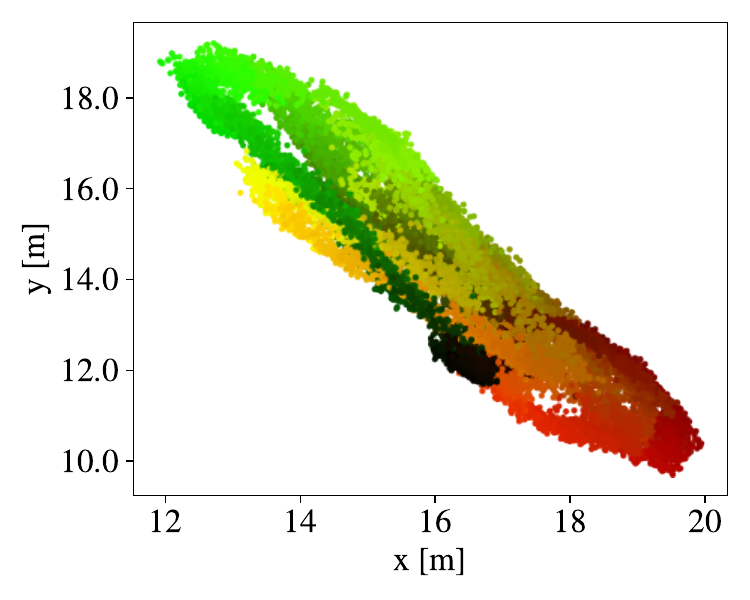}%
	\end{subfigure}%
    %\vspace{2pt}%
    
 	\begin{subfigure}{.49\columnwidth}%
		\includegraphics[clip, width=\columnwidth]{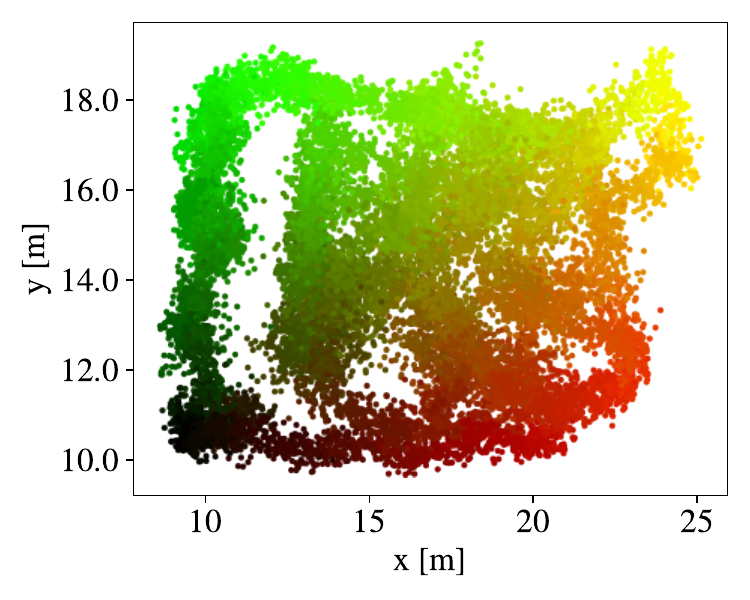}%
	\end{subfigure}%
    \hspace{2pt}%
 	\begin{subfigure}{.49\columnwidth}%
		\includegraphics[clip, width=\columnwidth]{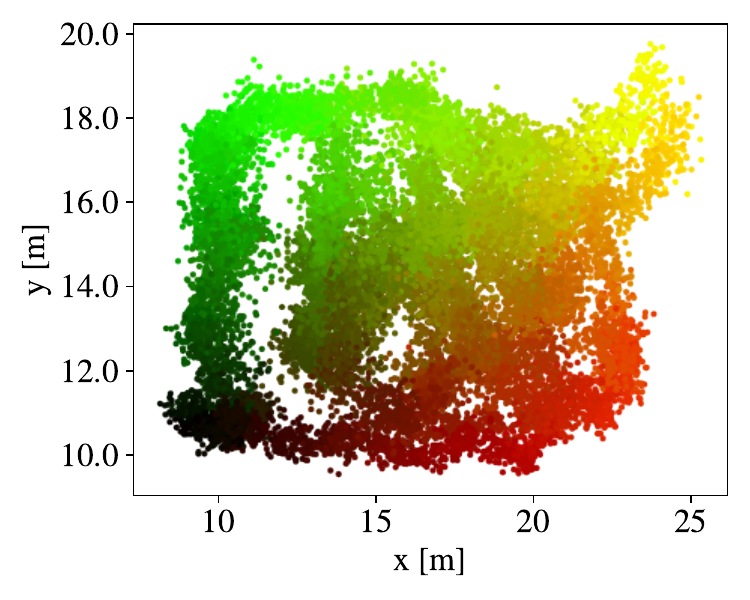}%
	\end{subfigure}%
		\caption{Channel charts for the 5G dataset generated with different time windows. In the top left corner, the ground truth trajectory is shown of the test dataset. In the other tiles, the channel chart for 5\,s (top right), 15\,s (bottom left) and 60\,s (bottom right) are shown.}
	\label{fig:res_5G_cc}
\end{figure}%

\begin{figure}[t]%
	\centering%
	\begin{subfigure}{.49\columnwidth}%
		\includegraphics[clip, width=\columnwidth]{img/CC_MIMO_gt.pdf}%
	\end{subfigure}%
	\hspace{2pt}%
	\begin{subfigure}{.49\columnwidth}%
		\includegraphics[clip, width=\columnwidth]{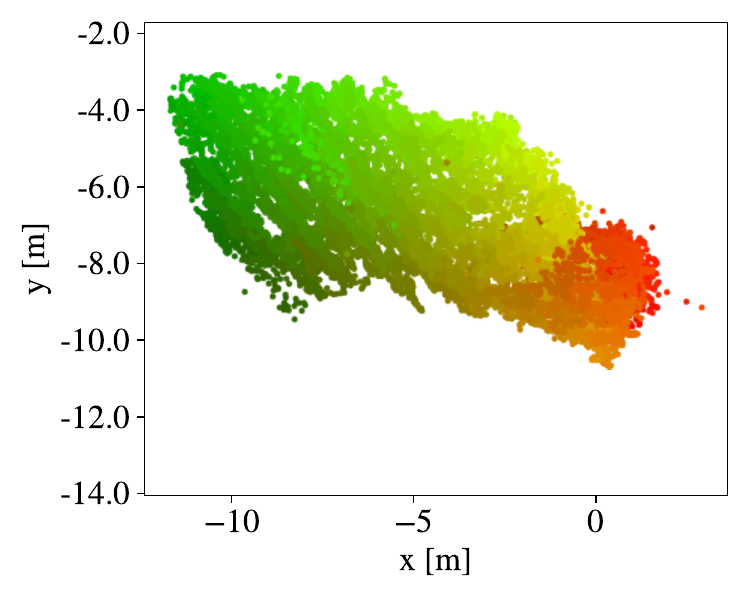}%
	\end{subfigure}%
    %\vspace{2pt}%
    
 	\begin{subfigure}{.49\columnwidth}%
		\includegraphics[clip, width=\columnwidth]{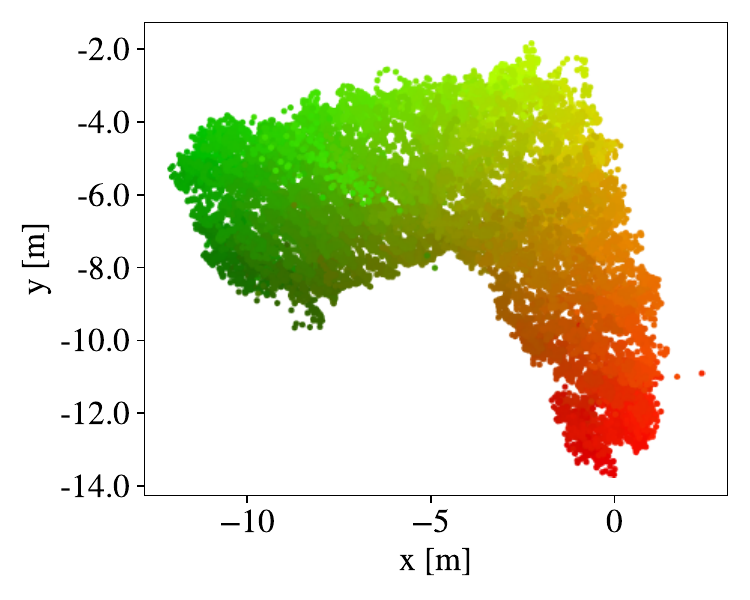}%
	\end{subfigure}%
    \hspace{2pt}%
 	\begin{subfigure}{.49\columnwidth}%
		\includegraphics[clip, width=\columnwidth]{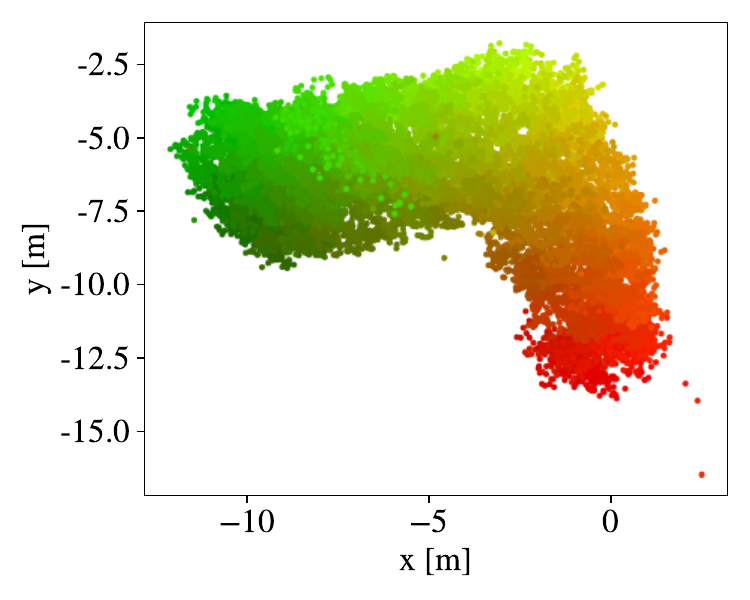}%
	\end{subfigure}%
	\caption{Channel charts for the SIMO dataset generated with different time windows. In the top left corner, the ground truth trajectory is shown of the test dataset. In the other tiles, the channel charts for 15\,s (top right), 30\,s (bottom left) and 60\,s (bottom right) are shown.}
	\label{fig:res_MIMO_cc}
\end{figure}

Fig.~\ref{fig:res_5G_cc} shows the ground truth of the test trajectory for the 5G test dataset, in the top left corner. 
In the other tiles, the channel charts are shown for 15\,s (top right), 30\,s (bottom left) and 60\,s right (bottom right). 
All channel charts show a good spatial consistency, while the 5\,s case does not recover the global geometry of the test data. 
The best results were achieved by the 15\,s window, on par to the results of supervised fingerprinting with a CE90 of about 0.87\,m for all noise levels for the channel chart and 0.90\,m for the fingerprinting. 
The results got worse with a window size of 60\,s with an increasing noise level from a CE90 of 0.84\,m for noise level 1 up to 1.11\,m for noise level 3. 
This is due to accumulating errors of the noisy velocity information. 
The larger the window, the larger the errors in the distance estimations, which leads to a worse channel-charting performance. 
However, the window size has to be sufficiently large to recover the global structure of the radio environment. 
We can see similar results in the SIMO evaluation shown in Fig.~\ref{fig:res_MIMO_cc}. 
The ground-truth trajectory (top left), is well recovered by the channel chart for the 30\,s window (bottom left). 
With noise level 1 with a CE90 of 0.55\,m, the accuracy of fingerprinting (0.56\,m) can be achieved, while the accuracy decreases with higher noise levels of up to 0.69\,m for noise level 4.

\subsection{Effects of varying numbers of BS}
\label{chap:numbers_BS}

This evaluation assesses the abilities of channel charting with different numbers of BS. 
The idea is to investigate the required spatial information contained in the input data to model the manifold only using a sparse distance matrix. 
We thus use a sufficiently large window to cover large distances within the trajectories. 
In the case of 5G, we used 15\,s as we have a mean velocity of 1\,m/s and thus have trajectories with the lengths of about 15m. 
For the SIMO dataset, we used a window size of 60\,s as the agent moves only about 0.28\,m/s on average, which leads to distances of about 17\,m within one window. 
The results are shown in Table~\ref{tab:res_number_bs}. 
As we use the TDoA values of the 5G radio system, described in Sec.~\ref{chap:preprocess}, to create the input tensors for the Siamese network, we need at least two basestations, while the SIMO radio system only consists of up to 4 basestations. 

\begin{table}[b!]
    \caption{Localization results for the channel charting (CC) compared to fingerprinting (FP) for different numbers of BS. The error, is the 90th percentile of the circular error (CE90). }
    \centering
    \setlength\tabcolsep{5pt}
    \begin{tabular}{cccccccccc}
        \toprule
        \multicolumn{2}{c}{} & \multicolumn{8}{c}{\# BS / Error (CE90 in m)} \\ \cmidrule{3-10}
        \multicolumn{2}{c}{} & 1 & 2 & 3 & 4 & 5 & 6 & 7 & 8 \\ 
        \midrule
        \multirow{2}{*}{5G} & CC & - & 5.31 & 1.26 & 1.21 & 1.23 & 0.97 & 0.92 & 0.82 \\
                            & FP & - & 2.67 & 1.28 & 1.12 & 1.38 & 1.05 & 0.99 & 0.90 \\
        %\midrule
        \multirow{2}{*}{SIMO} & CC & 1.23 & 0.63 & 0.56 & 3.25 &- & - &- & -  \\
                              & FP & 1.13 & 0.68 & 0.56 & 0.52 &- & - &- & -  \\
        \bottomrule
    \end{tabular}

    \label{tab:res_number_bs}
\end{table}

\begin{figure}[t]%
	\centering%
	\begin{subfigure}{.495\columnwidth}%
		\includegraphics[clip, width=\columnwidth]{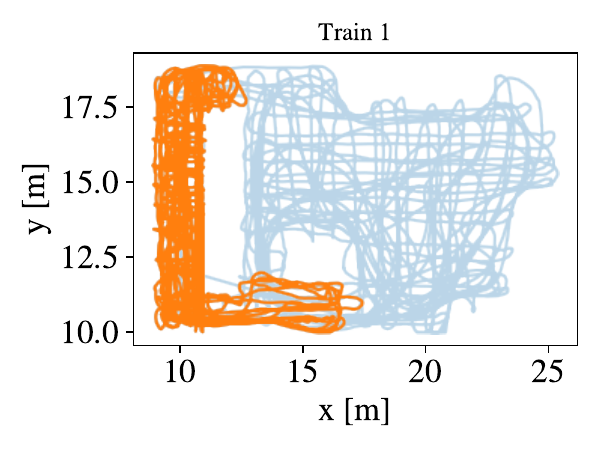}%
	\end{subfigure}%
	\hspace{2pt}%
	\begin{subfigure}{.495\columnwidth}%
		\includegraphics[clip, width=\columnwidth]{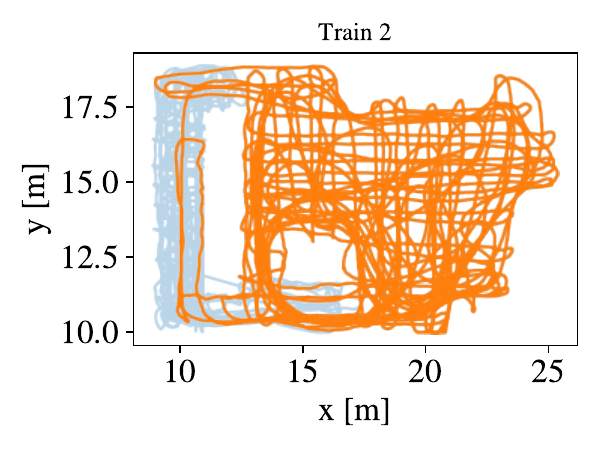}%
	\end{subfigure}%
	\caption{Trajectories of the training datasets, denoted in Table~\ref{tab:datasets}. The blue trajectory shows the combined datasets, while the trajectories of the specific datasets are shown in orange. The left plot shows the first training dataset and right shows the second.}%
	\label{fig:5G_train_data}
\end{figure}%

The results show that our velocity-based channel charting approach achieves competetive results to the fingerprinting model in most cases. 
In the 5G setup with only 2 BS, the localization accuracy is higher for fingerprinting compared to the CC, which indicates that the Siamese network may not have enough information available to model the manifold. 
However, with a higher number of BSs, the Siamese network successfully learned the manifold to be similar to the fingerprinting model. 
To train the 5G model, two different datasets were used with independent trajectories, shown in Fig.~\ref{fig:5G_train_data}. 
The combined trajectory is shown in blue, while the trajectories of the separate datasets is shown in orange. 
The first training dataset (left) has more datapoints on the left hand side of the area, while the second training dataset is more focused on the right area. 
Both datasets have no connection in the distance matrix. 
However, they overlap in some particular areas (bottom left), which means that they have similar CSI included in their trajectories. 
The Siamese network managed to combine both datasets into a single channel chart, which indicates that the Siamese network successfully learned the underlying manifold not only based on the distances provided by the velocity estimation but on the data itself.

\begin{figure}[t]%
	\centering%
	\begin{subfigure}{.495\columnwidth}%
		\includegraphics[clip, width=\columnwidth]{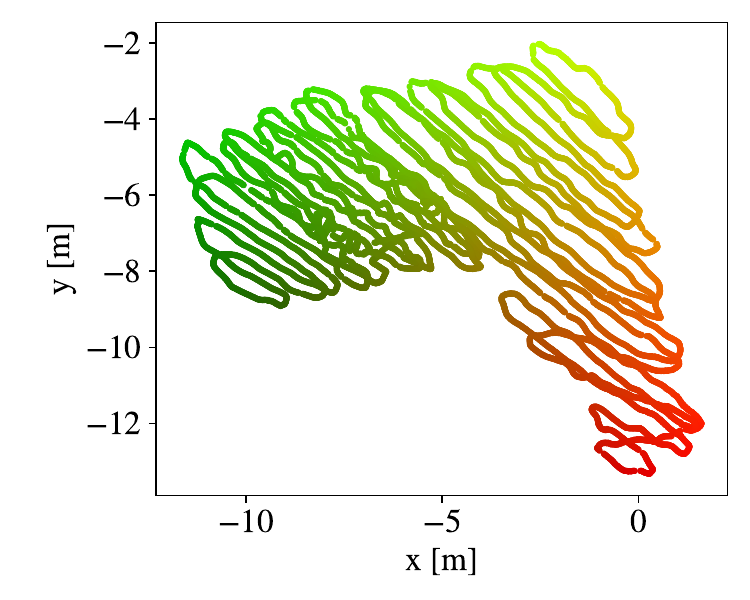}%
	\end{subfigure}%
	\hspace{2pt}%
	\begin{subfigure}{.495\columnwidth}%
		\includegraphics[clip, width=\columnwidth]{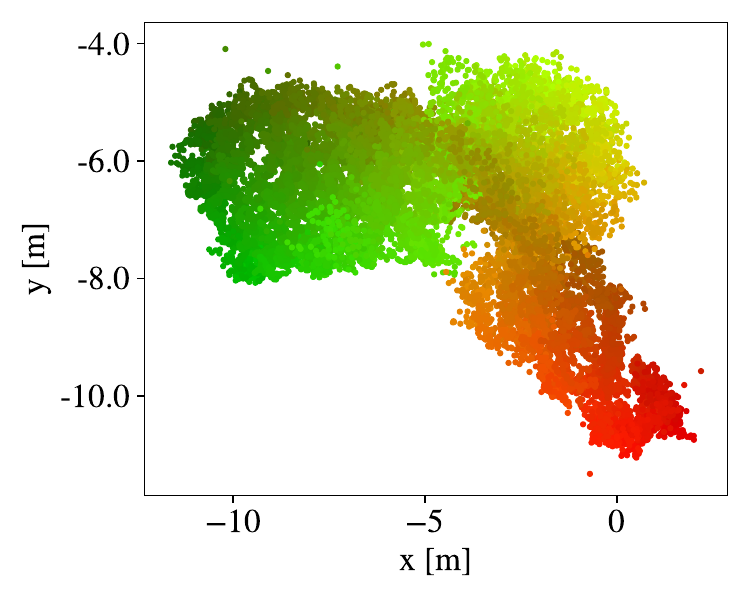}%
	\end{subfigure}%
	\caption{Channel chart with 4BS for the MIMO radio systems evaluated on the test dataset, shown on the left hand side. The estimated channel chart is shown on the right hand side. The color gradient shows the spatial consistency of the positions.}%
	\label{fig:res_MIMO_4BS}
\end{figure}

In the SIMO evaluation, the results are also similar to the fingerprinting model except the case with 4 BS. 
Fig.~\ref{fig:res_MIMO_4BS} shows the ground-truth locations on the left-hand side, while the channel chart is shown on the right. 
The channel chart still shows good spatial consistency, as the gradient is well recovered. 
However, in the area where $x<-5$, the channel chart appears to be twisted around the $x$-axis. 
We repeated the training for 50 times and found that only in 30 trials the Siamese network converged to the correct solution, while the other 20 trials failed similar to the chart shown in Fig.~\ref{fig:res_MIMO_4BS}. 
We calculated the mean of the Pearson correlation coefficients (PCCs) of the CSI for all BS combinations including all antennas of every BS, shown in Table~\ref{tab:SIMO_corr}. 
We can see that there are strong correlations between pairs of BSs, with a PCC of 0.86 for the combination BS1 and BS3 and 0.73 for the combination BS2 and BS4. 
While the BSs are very similar in the signal space, they are placed at different locations in the environment. 

We think due to this redundancy, the manifold is ambiguous and may have several representations in the 2D space. 
By comparing the results of the fingerprinting for 3BS and 4BS, we can also see no significant improvement. 
Hence, the 4th BS does not add any useful additional information to improve the fingerprint. 
By removing one BS, our Siamese network converged reliably, as we broke the symmetry and redundancy of the data. 
\begin{table}[t]
    \centering
    \caption{Mean Pearson correlation coefficients for every BS combination.}
    \begin{tabular}{ccccc}
    \toprule
            & BS1 & BS2 & BS3 & BS4 \\ \midrule
       BS1  &  1 & 0.61 & 0.86 & 0.62 \\
       BS2  &  0.61 & 1 & 0.65 & 0.73 \\
       BS3  &  0.86 & 0.65 & 1 & 0.63 \\
       BS4  &  0.62 & 0.73 & 0.63 & 1 \\
    \bottomrule
    \end{tabular}
    \label{tab:SIMO_corr}
\end{table}

\section{Limitations} \label{section:limitations}

Our experiments have shown that our velocity-based channel-charting algorithm can achieve accuracies similar to supervised fingerprinting, while being independent to radio topologies and architectures. 
However, the global consistency relies on the length of the trajectories considered in the distance matrix, which also means that the time windows have to be longer for agents with smaller velocities. 
For the 5G dataset, time windows of 15\,s were sufficiently large, while for the SIMO dataset at least 30\,s were required. 
We think that the trajectories need to cover a substantial part of the environment to achieve global consistency of the channel chart. 
Hence, the scaling to large environments is restricted. 
A possible solution for this could be to split the channel charts into distinct areas of limited size and switch between the models as shown in \cite{Stahlke2023Uncertainty}. 
However, as we can not spatially separate the recorded data, as we have no ground truth information, we may require external area identification. 
Another problem is that the error characteristics often do not depend on the movement speed of the agent like a bias in the angular velocity. 
Thus, slower agents may have more erroneous position estimations and hence distance errors, while faster agents may have better results with the same error characteristics. 
In consequence, the quality of the distance estimation and therefore that of the channel chart depends on the quality of the velocity-estimation system itself and on the movement pattern of the agent.

To transform the local channel-charting coordinates into the real-world environment, we employ topological map information, e.g., a floor plan, which only reflects the coarse geometry of the environment. 
While we can achieve superior results compared to the state of the art, the map-matching algorithm only works if a unique match of the channel chart in the map exists. 
This restricts the map matching to rotation-invariant and translation-invariant map information, and does not work for, e.g., rectangular areas without any unique features included like shelves or work desks. 
Additionally, the density and distribution of the channel chart is crucial for the map matching. 
While the map could provide a translation-invariant and rotation-invariant assignment due to unique features, the channel chart also requires a complete coverage of the areas around those features to avoid miss-alignment.

Our map-matching algorithm is also restricted to linear transformation with only rotation and translation. 
However, scaling errors due to, e.g., wrong step estimation as in the noise level 4 case, cannot be learned along with the probabilities of the map. 
The map-matching algorithm would scale the channel chart down and assign it to an arbitrary area, while learning that the surrounding areas do not have data assigned. 
Thus, before map matching is performed, the scale of the channel chart has to be estimated.

\section{Conclusion} \label{section:conclusion}

In this article, we proposed a framework for unsupervised fingerprinting only requiring velocity estimation and topological map information. 
Our velocity-based channel charting approach achieves accuracies of up to a CE90 of 1.16\,m for a 5G and 0.90\,m for a SIMO radio system, similar to supervised fingerprinting, even with very noisy velocity estimation and coarse map information.
Hence, our approach is applicable for low-cost sensor systems like smartphone-based PDR or odometry of robot platforms in combination with CSI recordings. 
Our adaptive map matching enables the usage of topological map information like floor plans to learn a transformation of the local channel-charting coordinates to the real-world environment. 
In contrast to the state of the art, our map-matching algorithm only needs a coarse representation of the environment, as it learns to adapt the map while it aligns the channel chart to the real-world coordinates.

\section{Acknowledgement}

We thank Maximilian Kasparek, Andreas Eidloth, Jan Niklas Bauer and Mohamed Soliman for implementing the proof-of-concept 5G uplink-TDoA positioning setup and its software-defined-radio-based processing pipeline.

\bibliographystyle{IEEETranM} % modified IEEEtran style for proper arXiv bibliography
\bibliography{main}%

\end{document}